\documentclass[12pt]{article}
\usepackage{epsfig}

\newcommand{\bmat}{\left(\begin{array}}
\newcommand{\emat}{\end{array}\right)}

\def\yzero{\smash{\hbox{$y\kern-4pt\raise1pt\hbox{${}^\circ$}$}}}

\def\beq{\begin{equation}}
\def\eeq{\end{equation}}
\def\beqa{\begin{eqnarray}}
\def\eeqa{\end{eqnarray}}

\def\-{\hphantom{-}}

\def\s2{\frac{1}{\sqrt2}}

\def\beq{\begin{equation}}
\def\eeq{\end{equation}}
\def\beqa{\begin{eqnarray}}
\def\eeqa{\end{eqnarray}}

\def\IF{\relax{\rm I\kern-.18em F}}
\def\II{\relax{\rm I\kern-.18em I}}
\def\IP{\relax{\rm I\kern-.18em P}}
\def\IC{\relax\hbox{\kern.25em$\inbar\kern-.3em{\rm C}$}}
\def\IR{\relax{\rm I\kern-.18em R}}

\def\Dsl{\,\raise.15ex\hbox{/}\mkern-13.5mu D} 
\def\IZ{Z\kern-.4em  Z}

\def\tgb{\tan{\beta}}

\def\asusy{a^{\rm SUSY}_\mu}
\def\bsg{b\to s\gamma}
\def\bmumu{B_s^0\to\mu^+\mu^-}

\def\lsim{\raise0.3ex\hbox{$\;<$\kern-0.75em\raise-1.1ex\hbox{$\sim\;$}}}
\def\gsim{\raise0.3ex\hbox{$\;>$\kern-0.75em\raise-1.1ex\hbox{$\sim\;$}}}

\catcode`\@=11   
\newdimen\@rotdimen
\newbox\@rotbox  

\def\@vspec#1{\special{ps:#1}}
\def\@rotstart#1{\@vspec{gsave currentpoint currentpoint translate
   #1 neg exch neg exch translate}}
\def\@rotfinish{\@vspec{currentpoint grestore moveto}}
\def\@rotr#1{\@rotdimen=\ht#1\advance\@rotdimen by\dp#1%
   \hbox to\@rotdimen{\hskip\ht#1\vbox to\wd#1{\@rotstart{90 rotate}%
   \box#1\vss}\hss}\@rotfinish}
\def\@rotl#1{\@rotdimen=\ht#1\advance\@rotdimen by\dp#1%
   \hbox to\@rotdimen{\vbox to\wd#1{\vskip\wd#1\@rotstart{270 rotate}%
   \box#1\vss}\hss}\@rotfinish}%
\def\@rotu#1{\@rotdimen=\ht#1\advance\@rotdimen by\dp#1%
   \hbox to\wd#1{\hskip\wd#1\vbox to\@rotdimen{\vskip\@rotdimen
   \@rotstart{-1 dup scale}\box#1\vss}\hss}\@rotfinish}%
\def\@rotf#1{\hbox to\wd#1{\hskip\wd#1\@rotstart{-1 1 scale}%
   \box#1\hss}\@rotfinish}%
\def\rotate{\@ifnextchar[{\@rotate}{\@rotate[l]}}
\def\@rotate[#1]#2{\setbox\@rotbox=\hbox{#2}\@nameuse{@rot#1}\@rotbox}

\catcode`\@=12

\topmargin
-1.5cm
\textwidth
15.5cm
\textheight
23.5cm
\oddsidemargin
0.7cm
\evensidemargin
0.7cm

\begin{document}

\makeatletter
\@addtoreset{equation}{section}
\makeatother
\renewcommand{\theequation}{\thesection.\arabic{equation}}
\pagestyle{empty}
\rightline{ IFT-UAM/CSIC-08-27}
\vspace{0.5cm}
\begin{center}
\LARGE{Modulus-dominated SUSY-breaking soft terms}
\LARGE{in F-theory and their
 test at   LHC \\[10mm]}
\large{ L. Aparicio, D.G. Cerde\~no and L.E. Ib\'a\~nez \\[6mm]}
\small{
Departamento de F\'{\i}sica Te\'orica C-XI
and Instituto de F\'{\i}sica Te\'orica UAM-CSIC ,\\[-0.3em]
Universidad Aut\'onoma de Madrid,
Cantoblanco, 28049 Madrid, Spain 
\\[10mm]}
\small{\bf Abstract} \\[8mm]
\end{center}

\begin{center}
\begin{minipage}[h]{16.0cm}
We study the general patterns of SUSY-breaking soft terms arising
under the assumption 
of Kahler moduli dominated SUSY-breaking in string theory models.
Insisting that all  MSSM  gauginos get masses at leading order and
that the top Yukawa  coupling is of order the gauge coupling constant
identifies the  class of viable models. These are  models in which the
SM fields live either in the bulk or at the intersection of local sets
of Type IIB $D7$-branes or their F-theory 
relatives. General arguments allow us to compute the dependence of the 
Kahler metrics of MSSM  fields on the local Kahler modulus of the
brane configuration in the large moduli approximation.   
We illustrate  this study in the case of toroidal/orbifold 
orientifolds but discuss  how the findings generalize to the 
F-theory case which is more naturally compatible with coupling
unification. Only three types of   7-brane configurations are
possible, leading each of them 
to very constrained patterns of soft terms for the MSSM.
We study their consistency with radiative electroweak 
symmetry breaking and other phenomenological constraints.
We   find   that essentially only the configuration 
corresponding to intersecting 7-branes is compatible
with all present experimental  constraints and
the desired abundance of neutralino dark matter.
The obtained MSSM spectrum 
is very characteristic and  could be tested at LHC. 
We also study the  LHC reach for the  discovery of this 
type of SUSY particle spectra.

\end{minipage}
\end{center}
\newpage
\setcounter{page}{1}
\pagestyle{plain}
\renewcommand{\thefootnote}{\arabic{footnote}}
\setcounter{footnote}{0}


\section{Introduction}

In the very near future LHC will start operating and a window for new physics will be
open. Low energy supersymmetry is one of the most solid candidates for 
that new physics.  If indeed low energy SUSY exists, a plethora of SUSY particles 
(gluinos, squarks, sleptons etc.) will be produced at LHC and their mass 
will be measured. Discovering SUSY at LHC would certainly be a  revolution.
On the other hand there is not at the moment a unique theory predicting the
values of masses for squarks, gluinos, sleptons...There are plenty of proposals 
for implementing the breaking of SUSY giving rise to different patterns 
of SUSY masses.  In any event,
if  low energy SUSY is found, measurements (however imprecise) of. e.g.  the 
value of squark and gluino masses would give very important information 
about the underlying theory. 

In the present article we  argue that a measurement of the 
SUSY spectrum of sparticles  would also provide an  experimental test  
for large classes of string compactifications giving rise to
standard model physics.  

In addressing these issues we will

\begin{itemize}
\item  1)  Assume that  the effective low-energy 
          $N=1$ supergravity approximation is  valid.

 \item  2) Assume that the low-energy theory has the structure of the MSSM.

\item   3) Assume that eventually all moduli are stabilized with a 
    vacuum energy  close to zero.

\item   4) Assume that SUSY-breaking predominantly originates in the vacuum expectation
       values  of auxiliary fields of the Kahler moduli $T_i$ of the
         compactification.

\item   5) Insist that all SM gauginos get a mass through the above mechanism at  leading order.

\item   6) Assume standard MSSM gauge coupling unification.

\item   7) Assume there is at least one  Yukawa coupling  (that of the top-quark) 
        of order $g$ (the gauge coupling constant).

\end{itemize}

Under these reasonable  assumptions one can identify a large class  of
string compactifications for which, in the leading  large volume 
approximation,  one can compute  the structure of SUSY-breaking soft terms
which in turn could be compared with data (if SUSY is actually found).

Let us briefly explain in more detail what underlies the 
above assumptions and what class of string compactifications they 
identify. The first two   conditions are just the expectation of being able 
to reproduce the MSSM  coupled  to  $N=1$ supergravity in a
perturbative regime (both in $\alpha$'
(string tension)  and string coupling $g_s$) 
from some successful string theory compactification.
The third assumption, that eventually all moduli gets dynamically fixed with
an almost vanishing cosmological constant, 
has been the subject of much work in
recent years \cite{kklt,reviewsfluxes}.
It has been shown that indeed the presence of Ramond-Ramond (RR)  and
Neveu-Schwarz (NS) antisymmetric fluxes in Type IIB orientifolds combined with
non-perturbative effects may stabilize all string moduli at weak coupling.
Specific examples have been worked out  (for reviews and references see \cite{reviewsfluxes}).

The fourth  assumption needs more explanation. It is well known that,
within the mentioned approximations the Kahler moduli in string compactifications have
a classical no-scale structure \cite{cfkn} 
in such a way that one can obtain SUSY breaking
with a vanishing (tree level) 
cosmological constant. This is true both in the Heterotic as well
as Type I, Type IIB  orientifold and  other compactifications and
corresponds to having non-vanishing vacuum expectation values for the
auxiliary fields of the Kahler moduli.
In Type IIB Calabi-Yau (CY) orientifolds this corresponds to
 string compactifications with  spontaneously
broken SUSY solving   the classical equations of motion
\cite{gkp}. 
It has also been shown \cite{gkp,ciu1,ciu2} that
 in those Type IIB orientifolds such SUSY-breaking corresponds to
the presence of RR and NS antisymmetric tensor fluxes
in the compactification. Such compactifications will {\it generically }  have
non-vanishing   fluxes so it looks like a most natural source of
supersymmetry breaking in
string theory. The scale of SUSY breaking may be hierarchically small
if the flux in the compact region where the SM sector resides is appropriately
small.

The 5-th  condition, asking for the presence of SUSY-breaking gaugino masses at
leading order  has a phenomenological motivation, as we will discuss in the main text.
It turns out to be quite restrictive.  Indeed, as we will argue, 
it excludes large classes of string compactifications
including perturbative heterotic compactifications, Type I string compactifications 
and Type IIB orientifolds with the SM residing on $D3$-branes at 
local singularities. In all these classes of compactifications 
the gauge kinetic function is independent of the (untwisted) Kahler moduli to leading order so that 
modulus dominance does not give rise to gaugino masses.
This argumentation  will leave us just with Type IIB orientifold compactifications 
with MSSM fields residing on $D7$-branes (or their intersections) and their
non-perturbative F-theory extensions.  
Alternatively one may consider 
the mirror description in terms of Type IIA orientifolds with the MSSM fields 
living at D6-branes   (or their intersections)  or  their non-perturbative extensions from M-theory
compactified on manifolds with $G_2$ holonomy. Being in principle equivalent 
we will  concentrate in the case of the MSSM chiral fields residing in the bulk 
of $D7$-branes or at the intersection of $D7$-branes (or F-theory 7-branes) since
at present little is known about the effective action of M-theory
compactifications in manifolds with $G_2$ holonomy (see \cite{acharya} for work
in this direction).

The 
unification of couplings is one of the most solid arguments in favour 
of the MSSM \cite{gcu}. We will thus  also  impose 
that gauge couplings of the MSSM unify
(6-th assumption)¡. A simple way in which this is achieved is
having at some level some  $SU(5)$, $SO(10)$ or $E_6$ symmetry relating the 
couplings. This unification structure, as recently emphasized in  \cite{dw,bhv},  may be
obtained most naturally in F-theory
\footnote{It is also naturally obtained in  heterotic
string compactifications. However in the perturbative heterotic case   
modulus dominated SUSY breaking does not give rise to
SUSY breaking soft terms to leading order, due to the
classical no-scale structure.}. It requires that the   
7-branes  associated to the SM gauge group all reside in the same stack.
Note that this does not imply the existence of an explicit GUT structure since
the symmetry may be directly broken to the SM at the string scale.
The value of the gauge coupling is related to the (inverse) size of the
4-volume  $V_7$ wrapped by the 7-branes. There is a  Kahler modulus whose 
real part $t$ is
proportional to this volume $V_7$. 
This is the local modulus which will be relevant for the
generation of the MSSM SUSY-breaking soft terms.
In particular 
we will  assume that a non-vanishing 
VEV for the auxiliary field of  this {\it local}  Kahler modulus $t$ 
exists giving rise 
to gaugino masses and other SUSY-breaking soft terms.

The final assumption  requires the existence of at least one Yukawa coupling
(that of the top quark) of order $g$ (the MSSM gauge coupling
constant), and has an 
obvious phenomenological 
motivation. It turns out to be quite restrictive too.
Chiral fields in $D7$-branes come in three classes depending on their 
geometric origin:  A-fields (from the vector multiplets A in the 
$D=8$ worldvolume action of the 7-brane and fermionic partners), $\phi$-fields
(from  $D=8$ scalar  $\phi $ multiplets and fermionic partners) and I-fields (fields from the intersection
of different 7-branes). As we will justify, this classification 
is valid both for 
simple (i.e. toroidal) Type IIB orientifold models as well as for
more general
F-theory configurations.  As we will see cubic Yukawa couplings of order one
may only be present for couplings of type (A-A-$\phi$), (I-I-A) or (I-I-I).
This substantially reduces the possible 7-brane configurations with a
potentially realistic phenomenology.

In order to compute the  SUSY-breaking scalar masses $m_\alpha $ and trilinear
couplings $A_{\alpha \beta \gamma}$ one needs to know the dependence of the matter metrics
of the MSSM chiral fields on the local Kahler modulus $t$. 
One can obtain this  $t$-dependence 
for the different type of fields $A,\phi, I$ which depends only on the local
geometry of the configuration of 7-branes. The results may explicitly be obtained
in simple toroidal orientifolds but one can argue  for the
validity of analogous results in the more general F-theory case.
One finds Kahler metrics depending like $K_\alpha =t^{-\xi{\alpha }}$ with $\xi_\alpha$ (the 
'modular weight' for the chiral field $\Phi_\alpha $)
 equal to $0,1/2,1$ for $\phi$,I, A respectively. 
We  also estimate subleading (in $\alpha $') corrections which
may arise from gauge magnetic fluxes  in the worldvolume of
7-branes (generically needed to get chirality) and argue that 
the Kahler metrics are flavour blind to leading order in $\alpha $'.
Armed with the knowledge of this  $t$-dependence for the Kahler metrics 
and the gauge kinetic function one can then explicitly compute the 
SUSY-breaking soft terms as a function of the auxiliary field of the
local Kahler modulus $t$. One can do that for the
three viable distributions  of matter fields consistent
with Yukawa couplings i.e.   (A-A-$\phi$), (I-I-A)  or (I-I-I). 
The obtained scalar soft terms are flavour blind to leading order,
which is good in order to avoid too large flavour changing neutral
currents (FCNC).

It is important to emphasize that
 our philosophy in this paper will be analogous to that in
e.g. \cite{bim1}. We will not assume a definite  mechanism which fixes
all moduli and then compute the soft terms but rather the reverse.
We will   rather assume as  above   that the auxiliary field
of a local Kahler modulus is the dominant source of SUSY-breaking
in the MSSM sector. Under these assumptions   we will end up
with very definite   patterns for the low energy SUSY spectrum.
If they are confirmed at LHC this would  tell us that the set of simple assumptions
here taken could be in the right track. This would also suggest that
one should concentrate in moduli fixing models leading to modulus
dominance in a IIB/F-theory setting (or dual models), gauge coupling unification and a
standard large  Fermi  scale/Planck scale hierarchy.
If they are not, a large class of
string compactifications with the above simplest  properties
would be ruled out. In any of both cases, if low energy SUSY is found,
 we will extract important
information about the underlying string theory compactification.

In the second part of this article we study the phenomenological
viability of these three options. To do this we run the soft terms
from the string/unification scale down to the electroweak (EW) scale
according to the renormalization group equations and impose radiative
breaking of the EW symmetry \cite{ir} in the standard way. We also
compute the full spectrum of sparticles and impose experimental
constraints, coming most notably from the bounds on the Higgs boson
mass, branching ratios of rare decays 
($b\rightarrow s\gamma$, $B_s^0\rightarrow \mu^+\mu^-$), 
and the muon anomalous magnetic moment $(g-2)_\mu$. 
Present limits on the masses of supersymmetric particles
are also imposed. We also comment on the existence of
dangerous charge and colour-breaking minima in the parameter space. 
Finally, the viability of the corresponding
lightest supersymmetric particle (LSP) as a dark matter candidate is
examined.

We present results for the phenomenological 
viability of the three string options and
find that the first ('non-intersecting') option with couplings (A-A-$\phi$)
is very constrained. The most severe bound is obtained by demanding the 
absence of tachyons in the scalar sector, 
which imposes $\tan\beta\leq 25$.
Also, this scenario fails to provide a viable dark matter candidate.
The LSP is always the stau, which is charged and should be unstable to be
cosmologically consistent. The simplest  way out in this case would be 
that R-parity would be broken with lepton violating dimension four couplings.
This would lead to very specific signatures at LHC.

On the other hand
the other two ('intersecting') options 
with couplings (I-I-A) or (I-I-I) give rise to similar   
results. However, obtaining the  appropriate  
amount of neutralino dark matter indicates a preference for  the (I-I-I) 
option in which both fermions and Higgs multiplets reside at
intersecting 7-branes.
Agreement with the WMAP results is obtained for $\tan\beta\simeq 40$.
 The spectrum of sparticles
may be accessible to LHC with a luminosity of $1-10$ fb$^{-1}$
by means of the missing energy signature.
The structure of the SUSY spectrum is quite specific and could also be tested at LHC.

The structure of this article is as follows. In the next section we study the
Kahler moduli dependence of the metrics of MSSM chiral fields both in IIB toroidal
orientifold models and more general 'swiss cheese' CY  and F-theory compactifications. We classify the
possible modular weights of matter fields and also the possible 7-brane
configurations consistent with the existence of trilinear couplings.
Subleading corrections to the metrics of matter fields coming from magnetic fluxes
in the worldvolume of 7-branes are also evaluated.
In section 3 the MSSM SUSY breaking soft terms are computed
for the different possible 7-brane configurations in terms of the
modular weights under the assumption of a single local Kahler modulus
domination. The low energy SUSY spectrum and radiative Electroweak (EW) symmetry
breaking is obtained in section 4 by numerically solving the renormalization
group equations from the string/GUT scale down to the EW scale. We examine the
experimental and dark matter constraints on the three different options for
MSSM fields at 7-brane configurations and discuss how the scenario with
all MSSM chiral fields at intersecting 7-branes can pass all tests.
We describe in section 5 the possible implications for LHC physics, in
particular 
the LHC reach for detecting squark and gluinos in this scheme through the
missing energy signature. Section 6 is left for our final comments and
conclusions.

\section{ Effective action of Type IIB orientifold models} 

As we mentioned, our assumption of Kahler modulus dominance 
for SUSY breaking combined with the requirement of 
non-vanishing leading order gaugino masses narrows very much 
the possible string compactifications. 
Present collider bounds on the mass of gluinos ${\tilde g}$  
imply $m_{\tilde g} > 195$ 
GeV at 95\% c.l. \cite{pdg}. Such large value for gaugino masses is difficult to
understand if the origin of gaugino masses were loop or other
type of subleading effects.  
Even taking into account the low-energy running of the gluino mass, such
large values seem to require the existence of a large unsuppressed
tree level  
contribution to gaugino masses.

 This seems to imply that only
models in which the MSSM gauge kinetic function depends
on the Kahler moduli at tree level are viable. This would exclude
Type IIB orientifold models with the MSSM fields residing
at either $D3$ or $D9$-branes. In both cases the gauge kinetic function
is given by the universal dilaton $S$-field
\footnote{See footnotes 3  and 6 for some qualifying  remarks 
to this point.}. It also excludes
perturbative heterotic vacua, whose gauge kinetic functions 
have the same property to leading order. Of course we do not
claim by any means that those widely studied kind of
vacua are not phenomenologically viable. We just concentrate in
the subset of string vacua in which the simplest assumption
of Kahler dominance gives consistent soft terms to
leading order.   
We are just left in the Type IIB case with orientifolds with
SM fields residing on $D7$-branes 
\footnote{Equivalent T-dual models may also be obtained from
IIB orientifolds with appropriately magnetized $D9$-branes
playing the role of $D7$-branes.
Those models would have an equivalent effective action and
will not be discussed here separately.}  
  and their 
non-perturbative F-theory generalizations with SM fields at
F-theory 7-branes. We will mostly concentrate in these classes of vacua
in what follows.

Before turning to that class of vacua we should emphasize again that
there are other dual vacua which would be equivalent 
from the point of view of the effective low-energy action
and hence we will not discuss these explicitly. In particular
that would be the case of Type IIB orientifolds with 
the SM fields residing at D5-branes or the S-dual
corresponding to non-perturbative heterotic vacua 
with the SM living on NS 5-branes. In both cases the 
gauge kinetic function depends on the Kahler moduli at leading order.
This is also the case for Type IIA orientifolds with
the SM fields residing at intersecting D6-branes or
coisotropic D8-branes \cite{coiso} which are mirror to the mentioned
Type IIB orientifolds with $D7$-branes. The same (mirror) effective
action is obtained although in this case the
role of the Kahler moduli is played by the complex structure
fields.

\subsection{Generalities}

We will concentrate in this section in the case of $N=1$ 
Type IIB orientifold
Calabi-Yau  compactifications
\footnote{For reviews on orientifold models see  \cite{interev} .}.
 In this class of models one projects
the string spectrum  
over $\Omega P$ invariant states, $\Omega$ being  the worldsheet 
parity operator and $P$ some particular order-2 geometric action 
on the compact CY coordinates. That operation leaves invariant certain 
submanifolds of the compact CY variety corresponding to
orientifold planes. We further concentrate on the case in which 
these orientifold planes are $O(3)$ and $O(7)$ planes.
In the first case the $O(3)$ planes are volume filling 
(i.e. contain Minkowski space) and are 
pointlike on compact space. The $O(7)$ planes will also
fill Minkowski space and wrap
orientifold invariant 4-cycles in the CY space.
Global consistency of the compactification will in general require the 
presence of volume filling $D3$- and $D7$-branes, the latter wrapping 
4-cycles $\Sigma_4$ in the CY. MSSM fields will appear from
open strings exchanged among these D-branes.

The closed string spectrum will include complex scalars, the axidilaton 
$S$, Kahler moduli $T_i$ 
and complex structure $U_n$ fields defined as 
\beq
S\ =\ \frac {1}{g_s}\ +\ iC_0 \ \ ;
T_j\ =\ \frac {1}{g_s (\alpha ')^2}{\rm Vol}(\Sigma_j)\ +\ iC_4^j \ ;\ U_n\ =\ \int_{\sigma_n} \Omega \,,
\label{losmoduli}
\eeq
where $g_s$ is the $D=10$ IIB dilaton, $\Sigma(\sigma)$ are 4(3)-cycles
in the CY 
and $\Omega $ is the holomorphic CY 3-form. Here $C_p$, $p=0,4$  
are Ramond-Ramond (RR) scalars coming from the dimensional reduction of 
antisymmetric p-forms. In terms of these the $N=1$ 
supergravity Kahler potential
of the moduli is given by
\beq
K(S,T_i,U_n) \ =\ -\log(S+S^*)\ -\ 2\log({\rm Vol}[{\rm CY}])\ -\ 
\log[-i\int \Omega \wedge {\overline \Omega}]\,.
\label{elkaler}
\eeq
To get intuition it 
is useful to 
consider the case of the 
(diagonal) moduli in a purely toroidal $T^2\times T^2\times T^2$ 
orientifold.
There are 3 diagonal $T_i$ Kahler moduli 
with Re$T_i=\frac {1}{g_s(\alpha ')^2} A_jA_k$, $i\not=j\not=k$,
$A_i$ being the area of the i-th 2-torus. The complex structure 
moduli $U_n$, n=1,2,3  coincide with the three  geometric  
complex structure moduli $\tau_i$ of the three  2-tori.
One then obtains
\beq
K\ =\ -\log(S+S^*) \ -\ \log(\Pi_i (T_i+T_i^*)) \ -\log (\Pi_n
(\tau_n+\tau_n^*))\,. 
\eeq
The reader familiar with $N=1$, $D=4$ string effective actions will
recognize the standard log (no-scale) structure of the
Kahler potential.
The gauge interactions and matter fields to be identified with the SM
reside at the $D7$, $D3$-branes and/or their intersections. 
The gauge kinetic function of the different
gauge groups may be obtained from the expansion of the 
Dirac-Born-Infeld action of the corresponding brane.  For a $D3$-brane and  a $D7$
brane wrapping a 4-cycle $\Sigma_j$ one gets  the simple result 
(see e.g. \cite{imr} for a derivation making use of dualities)
\beq 
f_{D7^j} \ =\ T_j\ \ \ \ \ ;\ \ \ \ f_{D3} \ =\ S\,.
\eeq
This difference is important: in modulus dominance the $D7$-brane
gauginos will get mass but those in  $D3$-branes will remain massless
\footnote{In terms of fluxes imaginary selfdual (ISD) fluxes give masses
to $D7$ but not to $D3$ gauginos, see e.g. \cite{ciu1,ciu2}.}.
As we emphasized,
this makes it phenomenologically problematic to build models with the
SM residing at $D3$-branes, as long as SUSY-breaking originates on the
Kahler moduli 
\footnote{In the case of $D7$-branes the gauge kinetic function can get $S$-dependent
corrections  if there are gauge fluxes in their worldvolume. For large Kahler moduli
the fluxes are diluted and the correction suppressed, see the discussion below.
On the other hand e.g. for $D3$-branes sitting at singularities the gauge kinetic
function does depend on the blowing-up Kahler moduli associated to  the singularity.
However it is difficult to make this compatible with gauge coupling unification
unless one stays in the singular limit. Furthermore, to assume that SUSY-breaking is dominated by
the blowing-up Kahler moduli would lead to highly non-universal soft masses, leading generically
to large FCNC transitions.}
.

\begin{figure}
\epsfysize=6cm
\begin{center}
\leavevmode
\epsffile{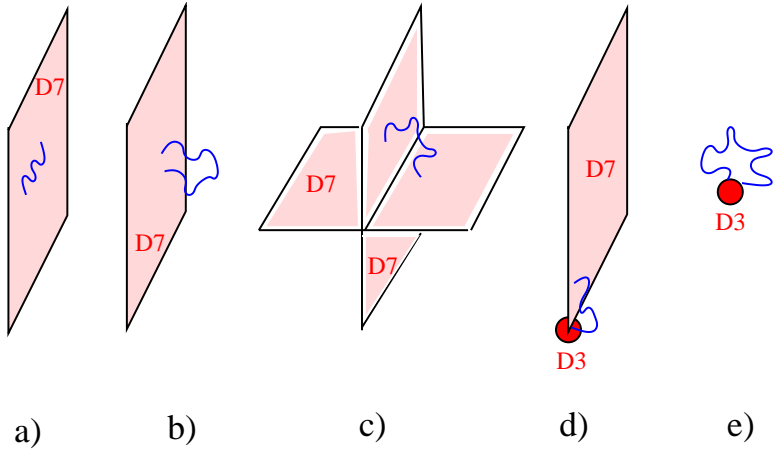}
\end{center}
\caption{Different origin of matter fields in $D7/D3$ configurations.
a) States from reduction of gauge fields A within the $D7$ worldvolume;
b) From reduction of D=8 fields $\phi$ parametrizing the $D7$ position;
c) From two intersecting $D7$-branes; d) From open strings between a
$D3$ and a $D7$; e) From open strings starting and ending on $D3$ branes.}
\label{soft1}
\end{figure}

In order to get a semirealistic model we need D-brane configurations 
giving rise to chiral gauge theories.
In the classes of IIB models here considered
one can classify the origin of chiral matter fields in terms of five
possibilities which are pictorially summarized in fig.\,\ref{soft1}.

The worldvolume theory of $D7$-branes is 8-dimensional  
supersymmetric Yang-Mills and contains both vector  A and matter
scalar  $\phi$ multiplets (plus fermionic partners) 
before compactification to $D=4$.
The first class of matter fields a) come from massless modes of the
gauge multiplet fields inside the $D7$ worldvolume (plus fermionic partners).
The case  b) corresponds to massless modes coming from the
$D=8$ scalars $\phi$  which parametrize the position of $D7$-branes
in transverse space (plus fermionic partners).
 The case c) corresponds to massless fields
coming from the exchange of open strings between intersecting
$D7$-branes. Open strings between $D3$ and $D7$ fields give rise to
matter fields of type d) whereas matter fields living fully on $D3$
branes correspond to type e).
Generically in order for  those zero modes to be chiral in $D=4$
additional ingredients are required. In particular there should be 
some non-vanishing magnetic field in the worldvolume of the 
$D7$-branes. In the case of $D3$-branes one may obtain chirality 
if they are located at some (e.g. orbifold) singularity in
the compact CY space.

In principle SM fields could come from any of these five configurations but, as we will see, there
are a number of constraints which reduce considerably the possibilities.
For the moment let us state that there are specific semirealistic constructions in which
SM fields live in any of these five configurations. For example, in section 9.1 of
\cite{ciu2} there are examples in which SM particles reside in the a,b and d
type of open strings above. Models with SM particles in $(33)$ branes 
at singularities
have been discussed  in \cite{aiqu,bjl,vw}. Concerning the c)   configurations,
they appear in magnetized $D7$-brane models which are mirror to IIA intersecting
D6-brane models \cite{cim1,cim2,ms,coiso,interev}

\subsection{Classification of 'modular weights'}

An important role in the computation of SUSY-breaking soft terms is
played by the  metric  $K_{\alpha \beta} $ of the  SM  matter fields $C_\alpha$,
which gives their  normalization. This metric is
in general a non-holomorphic function of the moduli
$K_{\alpha \beta }(S,S^*,T_i,T_i^*,U_n,U_n^*)$. 
Since we are interested in the case of modulus dominance 
we are particularly interested in the Kahler moduli dependence
of these metrics. Such metrics have been computed in a number of
string compactifications both in Heterotic and Type II cases and
for large volume  (i.e. to leading order in $\alpha '$) have   
a Kahler moduli dependence of the form 
\beq
K_{\alpha \beta } \ = \  \frac {\delta_{\alpha
    \beta}}{\Pi_i(T_i+T_i^*)^{\xi_i} }\,, 
\eeq
where $\xi_i$ are often called 'modular weights' following the 
heterotic terminology \cite{il} and are in general non-negative  rational numbers.
The precise dependence on the moduli depends on the geometry and
the origin of each field. We would like to discuss now the 
structure of these Kahler metrics in Type IIB orientifold 
models as well as their non-perturbative F-theory extensions.

\subsubsection{Kahler metrics in toroidal IIB orientifolds}

To be more explicit concerning the metrics it is useful to consider first the 
case of a $Z_2\times Z_2$  projection of a  Type IIB orientifold
of a factorized 6-torus $T^2\times T^2\times T^2$. The orientifold
operation is $\Omega (-1)^{F_L}R_1R_2R_3$ with $R_i$ being a reflection 
on the i-th complex plane. 
Tadpole cancellation conditions require the presence of
16 $D3$ branes and 3 different sets of 16 $D7^i$ branes, each wrapping a 
4-torus transverse to the i-th complex dimension.
As we mentioned, a variation of this orientifold with magnetic fluxes
in the worldvolume of some of the $D7$ branes was used to construct 
explicit semirealistic models \cite{ms,cim1,cim2}.
A  T-dual of this orientifold 
 model with $D9$ and 
$D5$-branes (without magnetic fluxes)
was first constructed in \cite{bl} and the structure of
Kahler metrics for the different matter fields was given in \cite{imr}
using duality arguments and in \cite{lmrs} through an explicit 
string computation (see also \cite{kn} and \cite{bblmr}).

The five type of matter fields discussed above correspond 
in the toroidal/orbifold case to states with 
a brane structure
\beq
a=(7^i7^i)_j \ ;\ b=(7^i7^i)_i \ ;\ c=(7^i7^j)  \ ; \ d=(37^i) \ ;\
e=(33)_i \ \ \ \ , \ \ \ i\not=j\,,  
\eeq
where 
the superindices label the three types of $D7^i$-branes  
 whereas the subindex in the 1-st,  2-nd and
5-th   cases correspond to the complex planes $j=1,2,3$.

As discussed in \cite{imr,lmrs}
these matter fields have the following Kahler metrics respectively:
\beq
K_{(7^i7^i)_j}= \frac {1}{t_k} \ ;\
K_{(7^i7^i)_i}= \frac {1}{s} \ ;\
K_{(7^i7^j)}= \frac {1}{s^{1/2}t_k^{1/2}} \ ;\
K_{(37^i)}= \frac {1}{t_j^{1/2}t_k^{1/2}} \ ;\
K_{(33)_i}= \frac {1}{t_i}\,.
\eeq
Here $s$ and $t_i$ are (twice) the real part of the complex dilaton $S$ and
Kahler moduli $T_i$ (i=1,2,3). Note that we have not included here the dependence
on the complex structure fields $U_n$, which will play no role
in modulus dominated SUSY-breaking. Note also that all metrics are flavour diagonal.
It is worth remarking the difference between the modular weights of the 
fields of type a) and b). This may be easily understood from the
point of view of the 8-dimensional $D7$-brane action. 
As we said the first class of matter fields a) come from massless modes of the
$D=8$ gauge multiplet fields A inside the $D7$ worldvolume.
The case  b) corresponds to massless modes coming from the
$D=8$ scalars  $\phi$  which parametrize the position of $D7$-branes
in transverse space. The dimensional reduction of this 8-dimensional 
action  gives rise to kinetic terms for  scalars $A_i$ inside the 
vector multiplet and scalars $\phi$ 
\beq
(\partial_\mu  A_i)(\partial^\mu A_j)\ g^{ij} \ \ +\ \ 
(\partial_\mu \phi)^2 \,, 
\eeq
where $g^{ij}$ is the inverse metric corresponding to the 
wrapped 4-cycle. The latter scales like $g^{ii}\simeq 1/R^4\simeq 1/t$
and gives the announced $t$-dependence for  the metric of those
fields. On the other hand the kinetic term of the scalar
involves no internal contraction and is hence t-independent.
Note that the massless modes from $\phi$  are in
general not particles transforming in the adjoint of the 
$D7$ gauge group as sometimes  stated in the literature.
They are in general chiral and may accommodate SM fields
(see e.g. the examples in \cite{ciu2}).

It is worth comparing this structure with that of perturbative heterotic
orbifold compactifications. In that case there is no dependence on the heterotic
axidilaton $S$ (to leading order) and  the modular weights depend on the orbifold symmetry and 
are  0,1 for untwisted matter and  fractional numbers multiples of  $(N-1)/N$ for
a $Z_N$ twisted sector \cite{il}. This is related to the fact that, being a closed string theory,
there are matter fields which belong to $Z_N$ twisted sectors.
Nevertheless the modular weights corresponding to the overall 
$T$-modulus is always integer.  On the
other hand in  a Type II orientifold matter fields correspond to
open strings which are never twisted. The only twist  possible for
open strings is that coming from  strings starting(ending) on Dirichlet 
boundary conditions and ending(starting)  on Neumann boundary conditions. 
This gives rise to an effective twist of order 2 leading to  
the presence of  possible modular weights 1/2 
in the above Kahler metrics
\footnote{Note that the presence of $s$ on the Kahler metrics makes  that the dependence 
on the $D=10$ dilaton $g_s$ through the definitions 
(\ref{losmoduli}) does not involve fractional powers.
This possible fractional dependence is  absent in the heterotic case 
since then, unlike the Type IIB case,  the 
supergravity Kahler moduli do not depend
on the $D=10$ dilaton.}.

In terms of the overall  Kahler modulus $t=t_i$
we will thus have  for the gauge kinetic function and
the chiral field metrics in this toroidal case :
\beq
f_a \ =\ T \ \ \ ;\ \ \ K\ =\ \frac {1}{s^{1-\xi}t^{\xi}}
\ ;\  \ \xi =0,1/2,1  \ .
\eeq
Fields of type a), d) and e) have $\xi=1$, those of type c) have
$\xi=1/2$ and type b) has $\xi=0$.

\subsubsection{Modular weight possibilities and Yukawa couplings }

One could then be tempted to conclude that the SM fields may  have any of these 
modular weights $\xi$ freely. 
This is not the case if the SM construction is to be realistic.
Indeed one important condition that one has to impose is that there 
should be at least  one Yukawa coupling of order 
the gauge coupling constant $g$, the one
giving mass to the top quark
\footnote{Masses for the rest of quarks and leptons, 
being much smaller,  could have their origin alternatively in non-renormalizable 
 Yukawa couplings and/or non-perturbative (e.g. string instanton) 
effects. That is extremely unlikely in the case of the top-quark.} 
. That means that 
{\it there should exist a trilinear superpotential } coupling
a Higgs field to the left and right-handed top quarks.
Now, the modular weights of fields with a trilinear coupling 
are not arbitrary. Consider again the 
 toroidal  $Z_2\times Z_2$ $N=1$ orbifold of \cite{bl}.
The possible Yukawa couplings among the five types of 
matter fields  may be easily understood in geometric terms 
and are classified in \cite{bl,imr}. One finds that 
Yukawa couplings of the following seven types exist:
\beqa
A)\  (7^17^2)(7^27^3)(7^37^1) \ \ [1/2,1/2,1/2] &  
B)\  (7^i7^i)_j(7^i7^k)(7^k7^i) \ \ [1,1/2,1/2] \\
C)\   (7^i7^i)_1(7^i7^i)_2(7^i7^i)_3  [0,1,1] &
D)\ (7^i7^i)_i(7^i3)(37^i) \ \ [0,1,1] \\
 E)   (7^i7^j)(37^i)(7^j3) \ \ [1/2,1,1] & 
 F)\    (33)_i(37^i)(7^i3) \ \  [1,1,1]  \\
G)\  (33)_1(33)_2(33)_3 \ \ [1,1,1] &  &
\label{sieteson}
\eeqa 
Again subindices label the three complex planes
and superindices the three types of $D7^i$-branes in this 
toroidal setting.
In square brackets are shown the values of $\xi$ for each of the three 
fields involved.
One observes that many possibilities for modular weights 
(like [0,0,0], [0,0,1], etc.) are not possible. 
In fact this is not just a property of the described 
toroidal setting but, as we argue below, it also applies to
more general CY IIB orientifolds and F-theory extensions.

Now, since the physical top-quark Yukawa coupling is large, it cannot have a 
non-renormalizable or non-perturbative origin. It should come from one
of the seven tree level possibilities above. Let us consider these options in turn.

\begin{itemize}

\item {\bf A)}
This corresponds to 3 sets of $D7$-branes, each pair overlapping over 
one complex compact dimension.
This gives rise to modular weights of type $(1/2,1/2,1/2)$.
Semirealistic Models of this class (with magnetic fluxes in some or all the branes)
have been constructed \cite{cim1,cim2,ms,coiso,interev}
. Those models are mirror to IIA orientifolds with
intersecting D6-branes.

\item {\bf B)}.
This case corresponds to states with couplings of type $(7^i7^i)_j(7^i7^k)(7^k7^i)$
with modular weights $(1,1/2,1/2)$.
Here one can identify  the first field with
the Higgs multiplet.
 Then quarks and leptons would come from overlapping
$D7$-branes as in the case A).

\item {\bf C),D)}

The  case C) corresponds to couplings $(7^i7^i)_1(7^i7^i)_2(7^i7^i)_3$
among open string states within the same stack of $D7$-branes.
They have modular weights $(1,1,0)$ (or permutations).
Note that in this  case in which the particles originate in the 
same stack of 7-branes there is a built-in asymmetry: two of the
states must have $\xi=1$ and the other $\xi=0$.
Then  it is natural to assign $\xi=0$ to the Higgs field
and $\xi=1$ to the squarks/sleptons.
Semirealistic models of this type
corresponding to $D7,D3$-branes at a $Z_3$ singularity may be constructed
(see e.g. section 9.1 of \cite{ciu2}). 
The case D), with similar overall modular weights,
correspond to models with both $D3$ and $D7$-branes. There are couplings
$(7^i7^i)_i(7^i3)(37^i)$ whose fields have analogous  Kahler metrics
 in terms of the overall Kahler modulus than the case C).
An example of these couplings are the leptonic couplings in the
same example of \cite{ciu2}. Quarks and leptons could live in the
$(7^i3),(37^i)$ open  strings and Higgs fields on $(7^i7^i)_i$.
Having the same modulus dependence, we will not consider this case D) as a
separate option in what follows.

\item{\bf E), F), G)}

These cases have in common that all or some of the SM gauge groups would have to reside on
$D3$-branes rather than $D7$-branes. Since we are all the time restricting ourselves to
modulus dominated SUSY-breaking, that means that the corresponding gauginos would be
massless to leading order. Thus, e.g., in the first case with a $(7^i7^j)(37^i)(7^j3)$
 coupling it is natural to locate the
QCD gauge group on $D3$-branes and that of the electroweak group on $D7$-branes. That would
imply having massless gluinos at this level. The opposite happens with
case F) with $(33)_i(37^i)(7^i3)$
couplings, which suggest locating the electroweak sector on the $D3$-branes.
Both situations are not viable phenomenologically.  Concerning the last case with
all SM fields on $D3$-branes, semirealistic models of this type
have been constructed in \cite{aiqu,vw,bjl}. In this case  all gauginos and scalars
would be  massless to leading order. This
yields equivalent results to modulus dominance in Heterotic compactifications.
Again this is  not viable phenomenologically unless we give up
our main assumption of modulus dominance and/or assume that all SUSY-breaking terms
come from subleading effects. The latter, although possible, is extremely
model dependent and would make the identification of any stringy
pattern on the MSSM SUSY-breaking spectrum quite difficult.

\end{itemize}

We thus conclude that under the assumption of an overall Kahler modulus
dominance,  there are essentially only three options, 
those corresponding to the
modular weight distributions $(1/2,1/2,1/2)$,
$(1/2,1/2,1)$  or $(1,1,0)$  with couplings of type A), B) and C) respectively.
From now on we will denote these three options by (I-I-I), (I-I-A)
and (A-A-$\phi$) respectively.

\subsection{Modular weights beyond the toroidal setting}

\begin{figure}
\epsfysize=4cm
\begin{center}
\leavevmode
\epsffile{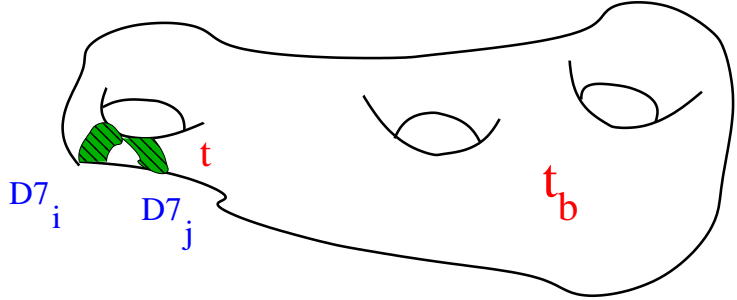}
\end{center}
\caption{  The SM $7$-branes wrap local 4-cycles of size $t$ in a CY whose
(larger) volume is controlled  by $t_b$.}
\label{instanton4}
\end{figure}

Toroidal  orientifolds are very special in some ways and we would like to see to
what extent the above findings generalize to more general IIB CY orientifolds. 
In particular $D7$-branes wrap 4-tori whose volume are directly related to the
overall volume of the compact manifold and then
the Kahler moduli appearing in the MSSM effective action are
directly related to the overall volume of the tori.
This is unnecessarily restrictive.
One would expect more general situations in which the
$D7$-branes wrap local cycles whose volume need not be directly related 
to the overall volume of the CY.
An example of this is provided by the 'swiss cheese' type of
compactifications discussed in \cite{conlon,ccq,bhp,acqs,ccq2}
(see also \cite{bmp}). In this class of 
more general models one  assumes that the SM resides 
 at stacks of $D7$-branes wrapping  'small cycles' in a CY whose overall
 volume is controlled by a large modulus $t_b$ (see fig.2) \cite{conlon}:
\beq
{\rm Vol}[{\rm CY}] \ =\ t_b^{3/2} \ -\ h(t_i)\,, 
\label{volcy}
\eeq
where $h$ is a homogeneous function of the 'small' Kahler moduli $t_i$ of degree 3/2.
The simplest example of CY with these characteristics is the manifold ${\bf P}^4
_{[1,1,1,6,9]}$ which has two Kahler moduli with \cite{ddf}
\footnote{For discussions for the multi Kahler moduli case see
\cite{bhp,ccq2,bmp}.}
\beq
{\rm Vol}[{\rm CY}] \ =\ t_b^{3/2} \ - \ t^{3/2}\,.
\label{dosmoduli}
\eeq
One important difference with the assumptions in \cite{conlon} is that we will not 
necessarily assume that $t_b$ is hierarchically larger than the $t_i$, rather we 
will just assume that $t_b$ is big enough so that an expansion in powers 
of $t_i/t_b$ makes sense. We will consider for simplicity the dependence on a
single small Kahler modulus $t$ and take both $t_b$ and $t$ sufficiently large
so that the supergravity approximation is valid.
 In this  case one can write a large volume expansion for the Kahler metrics
of the form  \cite{conlon,ccq}
\beq
K_{\alpha}\ = \frac {t^{(1-\xi_\alpha )}}{t_b} \,,  
\label{metrilla}
\eeq
with $\xi_\alpha$ the modular weights discussed above.
Note that for $t_b=t$ one would recover the metric we found in 
toroidal cases.
Although the explicit computation of the $t$-dependence in the
toroidal models is not directly applicable, one can use arguments based 
on the scaling of Yukawa couplings to indirectly compute 
the $\xi_\alpha$'s.
 The idea \cite{ccq} is to recall that the physical 
Yukawa couplings  ${\widehat Y}_{\alpha \beta \gamma}$ 
 among three fields $\Phi_\alpha $ after normalization is given in $N=1$ supergravity by
\beq {\widehat Y}_{\alpha \beta \gamma} \ =\ e^{K/2} \frac {Y^{(0)}_{\alpha \beta \gamma }}
{(K_\alpha K_\beta K_\gamma )^{1/2} }\,, 
\eeq
where $Y^{(0)}$ is the unnormalized Yukawa coupling which in 
Type IIB orientifolds is known to be holomorphic on the complex structure
fields $U_n$ and independent of the Kahler moduli $T_i$ (at least at the 
perturbative level we are considering)
\footnote{ This also justifies  a posteriori the $1/t_b$ dependence in eq.(\ref{metrilla}),
it is required to get a  holomorphic Yukawa $Y_{\alpha \beta \gamma }$
non dependent on the  large Kahler modulus $t_b$.}.  From that we conclude 
(using eqs.(\ref{volcy}-\ref{metrilla})) that 
the physical Yukawa coupling scales with $t$ like
\beq
{\widehat Y}_{\alpha \beta \gamma }\ \simeq \   t^{ 1/2 (\xi_\alpha + \xi_\beta +\xi_\gamma\ -\ 3)}\,.
\label{estay}
\eeq
Note that the dependence on the large modulus $t_b$ has dropped to leading order in 
$t/t_b$. This is expected since the wave functions of the matter fields are localized
at the small cycles and 
the Yukawa couplings are local quantities, 
expected to be independent   of the overall volume.
On the other hand one can estimate the $t$-scaling of the 
physical Yukawa couplings by noting that in Type IIB string 
theory those  may be found by computing
overlap integrals of the form
\beq
{\widehat Y}_{\alpha \beta \gamma }\ \simeq \ 
\int dx^6\ \Psi_\alpha \Psi_\beta \Psi_\gamma \,, 
\eeq 
where the integration is dominated by the overlap region of the three 
extra dimension zero mode states $\Psi$
\footnote{See \cite{cim2} for an explicit computation of Yukawa
couplings in a semirealistic D-brane model by the overlap
integral of three internal wave functions.}
. This assumes those internal
wave functions to be normalized:
\beq
\int |\Psi_\alpha|^2\ = \ \int |\Psi_\beta|^2\ =\ \int
|\Psi_\gamma|^2\ =\ 1\,.
\label{normal}
\eeq
Consider for example the case of three sets of $D7_i$ branes i=1,2,3, 
each overlapping with each other on  a complex curve of real dimension two.
This would be analogous to the case  A)  in the toroidal discussion above.  Then
the matter fields at each intersection overlap over a 2-dimensional space 
and, given eq.(\ref{normal}),  their wave function would scale like $1/R\simeq t^{-1/4}$.
For a trilinear coupling to exist the three $D7$-branes overlap
in this case 
at a point so that the  physical Yukawa should scale like 
${\widehat Y}\simeq   (1/t^{1/4})^3 = t^{-3/4}$.  This agrees with (\ref{estay}) 
if the three chiral fields at intersecting $D7$-branes have $\xi_\alpha=1/2$.
Using  a similar argument, this time using the Yukawa coupling 
of type (I-I-A) one finds that bulk fields of type A have $\xi_\alpha =1$.
Finally from the existence of couplings of type $A-A-\phi$ one
gets that the zero modes from fields of type $\phi $ have $\xi_\alpha =0$.
Note that for $t_b\propto t$ the Kahler metrics from 
(\ref{metrilla}) will then have the same $t$-scaling that we found
for the different type of zero modes A, $\phi$,I in the toroidal case.

The above results are rather general and independent 
of the particular geometry as long as one can approximate the scaling behaviour
with a single local Kahler modulus. Note also that one can only extract information on the
sum of the three modular weights (not the individual ones) and obviously
only in models in which those trilinear couplings actually exist

\subsection{F-theory, coupling unification and  modular weights.}

Our discussion in the previous section applied to general Type IIB CY orientifolds 
in which the SM fields reside at $D7$-branes wrapping 'small' cycles characterized 
by an overall local modulus $t$. Prototype models in this class would  include 
intersecting $D7$-brane models in which the gauge group factors 
$SU(3)$, $SU(2)$ and $U(1)$ of the SM 
live on different stacks of branes. In this case 
each gauge coupling is independent and  
 the observed joining of the MSSM 
coupling constant at a scale of order $10^{16}$ GeV would be a coincidence.
It would be interesting to have a Type IIB framework with the SM on 7-branes
in which gauge coupling unification could be a natural property from the
beginning. This  would also be more consistent with our 
approach which uses a single Kahler modulus $t$ to describe the 
Kahler metrics and gauge kinetic function.

It has recently been pointed out \cite{dw,bhv} that one can naturally achieve gauge 
coupling unification within a type IIB scheme by going to  F-theory 
non-perturbative constructions. This is not the place to give an
introduction to these  constructions but we will
discuss here briefly how our discussion above is  extended to
F-theory compactifications of the class introduced
in \cite{dw,bhv}. For an introduction to F-theory and 
references see the first paper in \cite{reviewsfluxes}.

F-theory \cite{ftheory} 
may be considered as a non-perturbative version of Type IIB
orientifolds. Type IIB string theory in $D=10$ has a 
non-perturbative   $SL(2,{\bf Z})$ S-duality 
symmetry under which the complex field  $\tau =\frac {1}{g_s}+iC_0$
transforms. Here $1/g_s$ is the Type IIB dilaton and $C_0$ is the  RR scalar. 
The idea in F-theory is to identify locally this $\tau$ with  the complex 
structure of a 2-torus (also having this 
$SL(2,{\bf Z})$  symmetry) living in extra 11-th and 12-th dimensions.
Thus F-theory gives a geometric description of the S-duality symmetry
in compactifications of Type IIB theory.  One considers compactifications
of this 12-dimensional theory on a CY complex 4-fold $M$  down to $D=4$.
The CY 4-fold must be elliptically fibered over a complex 3-dimensional CY $B$,
meaning that locally one can write $M=T^2\times B$ with the complex structure
modulus of the $T^2$ identified with $\tau$. These  are  clearly 
non-perturbative vacua since the $SL(2,{\bf Z})$ symmetry  
includes transformations under which $g_s\rightarrow 1/g_s$.
Still  the effective $D=4$, $N=1$ low-energy action 
will admit a standard perturbative supergravity description in the 
limit of large Kahler moduli.

In \cite{bhv} specific F-theory models with
7-brane content with GUT-like structure were discussed. These are 
{\it local} models in which only the local GUT physics 
associated to a chain of singularities is specified.
In these models  F-theory is compactified 
on complex  4-folds with the local structure 
\beq
M\ =\ T^2\times B\ =\ K3\times S\,, 
\eeq
 in which the $K3$ surface  
is itself elliptically fibered. The theory contains F-theory 
 7-branes which wrap the complex 2-fold $S$. Inside the 
3-fold B these 7-branes  correspond to 
complex codimension 1 singularities.
Depending on the canonical ADE classification of the singularities 
the gauge groups of 7-branes have   $A_n$, $D_{n}$ or exceptional
$E_{6,7,8}$ algebras. This corresponds to
gauge groups $SU(n+1)$, $SO(2n)$ and $E_6,E_7,E_8$. 
Thus the gauge group in F-theoretical 7-branes goes beyond 
what one can get  in perturbative Type IIB orientifold 
$D7$-branes in which
only $SU(n+1)$ and $SO(2n)$ gauge groups may be obtained. 
Furthermore in F-theory the matter content in models with
$SO(2n)$ gauge symmetry may include spinorial representations 
which are not present in perturbative IIB orientifold compactifications.
This is important since e.g. it allows for an underlying 
$SO(10)$ GUT structure which contains SM fields in 16-dimensional
spinorial representations.

The $D=4$ chiral matter fields in F-theory have 
the same qualitative origin as in perturbative IIB 
orientifolds. There are chiral fields of type  A and $\phi$ 
which originate from zero modes upon compactification 
of the $D=8$ 7-brane worldvolume gauge field $A$ and 
scalar  $\phi$ (plus fermionic partners). 
As in the perturbative case,
in  order to get chirality one needs to
add additional  magnetic backgrounds 
which break the original 7-brane gauge symmetry 
as $G\rightarrow H$.
 Then superpotential couplings of the schematic form
\beq
Y_{\alpha \beta \gamma}\ =\ \int_{S}\ h_{nlm}\ A^{\alpha n}A^{\beta l}\phi^{\gamma m}
\eeq
are in general present \cite{dw,bhv}. Here the coupling constant 
is evaluated as an overlap integral  of the three wave functions
over the 4-dimensional manifold $S$ which is wrapped by the 
corresponding 7-brane. 
The  $h_{nlm}$ are gauge structure constants in the branching $G\rightarrow H$.
This is analogous to the  couplings of type C) which
we discussed for the perturbative case. One interesting point noticed in
\cite{bhv} is that in fact for the simplest class of F-theory
compactifications (with S either Hirzebruch or del Pezzo 
surfaces) there are no $\phi$ zero modes and hence no 
Yukawa couplings are possible. This is important since it indicates 
that in F-theory Yukawa couplings of type (A-A-$\phi$) are
difficult to obtain and hence this class of models
is phenomenologically problematic.

In addition, chiral zero modes may appear at the
intersection of F-theory 7-branes over 
a complex Riemann curve $\Sigma$ inside $S$. 
In the F-theory language this corresponds to 
the case in which the rank of the singularity type of a 
7-brane can increase along this  complex curve $\Sigma$
(e.g. from $D_{5}$ to $E_6$).
 This is analogous to the 
case we mentioned for the toroidal setting with two 
$D7$ branes intersecting  over a common $\Sigma=T^2$.
The overlap takes place in 6 real dimensions (Minkowski$\times \Sigma$)
and hypermultiplet matter fields  $\Lambda,\Lambda^c$ 
appear coming from open strings going from one 7-brane with base $S$ to another 
with base $S$'. Each 7-brane carries its  own gauge group $G_S$ and $G_{S'}$
and the hypermultiplets transform as 'bifundamentals' with respect to the 
$G_S\times G_{S'}$ group. We talk of bifundamentals here in a generalized sense
since e.g. fields may transform like spinorials or antisymmetric tensors
with respect $G_S$ or $G_{S'}$ 
\footnote{Here $G_S$ and $G_{S'}$ need not be SM gauge factors, one could have 
e.g. $G_S=SO(10)$, $G_{S'}=U(1)$ with $G_S$ broken down to the SM with the addition
of  gauge backgrounds on the complex curve $\Sigma$.}. 
 In this case there are trilinear Yukawa couplings 
which involve
an overlap integral over the complex curve $\Sigma$ 
\beq
Y_{\alpha \beta \gamma}\ =\ \int_{\Sigma} \ c_{ijk}\ \Lambda^{c;\alpha i} A^{\beta j} \Lambda^{\gamma k}\,, 
\eeq
where the $A^{\beta j}$ are  zero modes originated from  $D=8$ gauge fields either in one 
7-brane or the other. Here $c_{ijk}$ are structure constants associated to the branching 
$G_{\Sigma}\rightarrow G_S\times G_{S'}$, where here $G_\Sigma$ is the group corresponding
to the enhanced singularity. 
Note that this is analogous to the Yukawa couplings of type (I-A-I) which we described in
the toroidal case. 

There is also the possibility of two 'matter curves'  $\Sigma_1$ and $\Sigma_2$ in the base
$S$  to collide in a point. This would correspond to a triple intersection  of 7-branes.
At the intersection of $\Sigma_1$ and $\Sigma_2$ the rank of the singularity type would
be increased in two units. An example in which a singularity of type $SO(12)$ 
is enhanced   to a $E_8$ type singularity is described in \cite{bhv}. Chirality is 
obtained by adding a $U(1)$ magnetic background in the underlying 7-brane 
with gauge group $SO(12)$. The effective low-energy group is $SO(10)$ 
with 3 generations of spinorial 16-plets ('bifundamentals' under the $E_8$ subgroup
$SO(12)\times U(1)_1\times U(1)_2$) living at two intersection matter curves $\Sigma_1$ and $\Sigma_2$.
The Higgs 10-plets arise from an additional matter curve $\Sigma_3$.
In these multiple 7-brane  intersections trilinear Yukawa couplings involving   
three 'bifundamentals' exist. In this coupling  wave functions  overlap in a point 
and one has a structure 
\beq
Y_{\alpha_j \beta_k \gamma_l} \ =\  f_{abc} 
\Lambda^{\alpha_j a}\Lambda^{\beta_k b}\Lambda^{\gamma_l c}\,.
\eeq
This is the F-theory analogue of the (I-I-I) couplings that we described in the
toroidal setting.

We thus see that the three classes of Yukawa couplings that we discussed for 
 perturbative Type IIB orientifolds are the ones which are present in the 
local F-theory non-perturbative models of \cite{dw,bhv}.
 Note that the Yukawa couplings are again obtained as overlap integrals
with support on spaces of real dimension 4, 2 and 0 respectively. Thus our arguments 
to extract the leading large Kahler modulus behaviour of the Kahler metrics of
matter fields go through in these  local F-theory models. We thus again have that, 
restricting ourselves to the dependence on an overall local Kahler 
modulus $t$ the matter fields of type  A, I and $\phi$ have respectively 
modular weight 1, 1/2 and 0.

In many aspects the results are quite analogous to the perturbative 
orientifold case in which gauge groups reside on standard $D7$s.
The main difference is that 'bifundamentals' in F-theory may contain
full SM generations (like in the spinorial of $SO(10)$) and the full
gauge group of the SM may reside in a single F-theory 7-brane stack.
Thus in these more general schemes gauge coupling unification 
is a natural consequence.

\subsection{Subleading effects from magnetic gauge fluxes}

In previous sections we have mentioned that getting chirality of the 
massless $D=4$ spectrum in general requires the presence of magnetic fluxes
\footnote{These open string gauge  fluxes should not
be confused with antisymmetric RR and NS fluxes which may break SUSY
and come from the closed string sector.} 
on the worldvolume of $D7$ or F-theory 7-branes. So a natural question is how
these fluxes may modify the Kahler metrics of matter fields 
and the gauge kinetic functions associated to 7-branes.
In  the large $t$ limit the magnetic fluxes are diluted and 
one expects this effect to be subleading. To flesh out this expectation 
we study in this subsection how the presence of magnetic fluxes affects the 
effective action in  the well studied case of toroidal/orbifold 
perturbative IIB orientifolds. The presence of fluxes affects in the same manner
perturbative and F-theory compactifications so one expects similar corrections to
appear in more general CY orientifolds or F-theory compactifications.

 In configurations with sets of overlapping 
$D7$-branes in generic 4-cycles,  chirality is obtained by the
addition of magnetic fluxes. 
In the case of toroidal/orbifold IIB orientifold models
the corrections to the Kahler metrics coming from 
magnetic fluxes have been computed in \cite{lmrs}.
The results for states coming from strings within 
$D7$s wrapping the same cycle is
\beq
K_{(7^i7^i)_j} \ =\ \frac {1}{t^k}\ |\frac{1+iF^k}{1+iF^j}| \ \ ;\ \ 
K_{(7^i7^i)_i} \ =\ \frac {1}{s}\ (1+|F^jF^k|)\,, 
\eeq
where $i\not=j\not=k$  label the 3 2-tori and $F^i$ is 
the magnetic flux going through the 
i-th 2-torus which may be written as
\beq
F^i \ =\ n^i  (\frac{s t_i}{t_jt_k})^{1/2} \,, 
\eeq
with $n^i$ quantized integer fluxes.
For states coming from  open strings in between 
 (magnetized) branes  $D7^a$, $D7^b$ wrapping different 
4-tori one has
\beq
K_{7^a7^b}\ =\ \frac {1}{(s t_1t_2t_3)^{1/4}}
\Pi_{j=1}^3 \ u_j^{-\theta_{ab}^j}\
\sqrt{ \frac {\Gamma (\theta_{ab}^j)} {\Gamma (1-\theta_{ab}^j) } }\,, 
\label{metrica}
\eeq
where $u_j$ are the real parts of the complex structure moduli,
$\Gamma $ is the Euler Gamma function and 
\beq
\theta _{ab}^j \ =\ \arctan(F_b^j)-\arctan(F_a^j) \ .
\eeq 
The gauge kinetic functions are also modified in the presence of
magnetic fluxes as
\beq
Ref^a_i \ =\  T_a \ ( 1 \ +\ |F_a^jF_a^k|) \ .  
\eeq

These results apply for the case of a toroidal and/or 
orbifold setting. However we can try to model out what could 
be the effect of fluxes in a more general setting following 
the above structure. To model out the possible effect of fluxes 
we consider again the 
limit with $t_i = t$ and diluted fluxes $|F_i|= F$, i.e. large $t$
and ignore the dependence on the complex structure $u_i$ fields.
Then from the above formulae one obtains  the dilute flux results 
\beqa
K_{(7^i7^i)_j}^i \ &=& \ \frac {1}{t}\  \ \ ;\ \
K_{(7^i7^i)_i} \ =\ \frac {1}{s}\ (1 + a_i \frac {s}{t})\ ,  \\
K_{7^a7^b}\ & =& \ \frac {1}{(s^{1/2}t^{1/2}) } (1\ +\  c_{ab}\ \frac
{s^{1/2}}{t^{1/2}})\ , \\
{\rm Re}f_i \ &=&\   t \ +\  a_i  s \ ,
\label{metricaaprox}
\eeqa
where $a_i,c_{ab}$ are constants (including the flux quanta) of order one.
The limit of eq.(\ref{metrica}) has been obtained by expanding the 
Gamma functions for dilute flux and ignoring the dependence through the
complex structure fields $u_i$
\footnote{As an example,  for the specific semirealistic model analyzed in 
detail in \cite{fi}
one has $F_1^2=F_1^3=3^2 (s/t)$, $F_2^i=F_3^j=0$ with 
 $a_1=3^2$, $c_{12}=c_{13}=3ln4$,  $a_2=a_3 =c_{23}=0$.}.
The above three formulae may be summarised by:
\beq 
K_{matter} \ =\ \frac {1}{s^{(1-\xi)}t^\xi } \times (1\ +\ c_\xi (s/t)^{1-\xi}) \ =
\ \frac {1}{s^{(1-\xi)}t^\xi} 
\ +\ \frac{c_\xi}{t}\, ,
\label{esta}
\eeq
with $c_{\xi}$ some flux-dependent constant coefficient whose value will 
depend on the modular weight $\xi$ and the magnetic quanta.
This means that the addition of fluxes has the effect in all cases to add a term which goes like $1/t$.
Heuristically,  we know that with the addition of large fluxes the $D7$-branes turn into localized branes,
very much like $D3$-branes and we know that matter fields for $D3$-branes have kinetic terms which go
like $1/t$. This would explain the modulus dependence of the last term
in eq.(\ref{esta}).

Note that in the diluted flux limit ($s/t\rightarrow 0$) one recovers
the metrics we discussed above. Note also that the metric for the
matter fields of type $(7^i7^i)_j, j\not=i$ does not get corrections 
to leading order. As we will see, this will imply that 
 fields with modular weight $\xi=1$ 
will remain massless even after SUSY-breaking. This will be phenomenologically
relevant, as we will discuss in the numerical analysis.

For a non-toroidal compactification one expects analogous flux corrections to
exist with $\xi =0,1/2,1$ corresponding to chiral fields of types
$\phi $, I and A respectively.
One can easily generalize this  to the case in which $t$ is the local
modulus coupling to the MSSM in a CY whose volume is controlled by a larger
modulus $t_b$, as we discussed in section (2.3). 
From the above expressions one  then finds a Kahler moduli
dependence of the matter metrics of the form
\beq
K_{matter} \ =\ \frac {t^{1-\xi}}{t_b} \ (1\ +\ c_\xi t^{\xi-1}) \ =\ 
\frac {t^{(1-\xi)}}{t_b} \ +\ \frac {c_\xi}{t_b} \ .
\label{flujillas}
\eeq
as a generalization of eq.(\ref{metrilla}).
We have ignored here the explicit dependence on $s$ and the complex structure,
not relevant for the computation of soft terms.
Note that the flux correction actually will depend on the large 
modulus $t_b$ rather than on the local modulus $t$.

An additional comment is in order concerning flavour dependence 
of the Kahler metrics.  In the three schemes with viable Yukawa couplings
described above all quarks and leptons have the same $t$-dependent Kahler metrics.
 Magnetic fluxes may introduce 
some flavour dependence
 but that will be suppressed in the dilute
limit here considered. 
On the other hand this is still compatible  with the existence of a 
non-trivial flavour dependence in the Yukawa coupling sector.
Indeed in Type IIB orientifolds the holomorphic Yukawa
(perturbative) superpotentials are independent from the Kahler moduli 
and depend on the complex structure fields.
Thus flavour dependence in these compactifications resides in the
complex structure sector. This fact has also been 
recently emphasized in \cite{ccq,mirror}.
 Since we are assuming that SUSY-breaking is 
taking  place only in the Kahler moduli sector, no
flavour dependence will appear to leading order 
in the Kahler metrics and, consequently, in the
obtained SUSY-breaking soft terms.

As a final observation, note that with SUSY-breaking 
controlled by the auxiliary field $F_t$ of a particular local modulus $t$,
the complex phases which might appear in soft terms can always be
rotated away \cite{fluxed}. So there will be no  SUSY-CP problem and
there will be no SUSY contribution to 
 the electric dipole moment of the neutron
coming from soft terms.

\section{Supersymmetry breaking and soft terms}

With the knowledge of 
  the matter metrics and gauge kinetic function 
we can now address the computation of SUSY-breaking soft terms.
We have argued that similar dependence for Kahler metrics on the local Kahler modulus $t$ 
is obtained both for perturbative IIB CY orientifolds and F-theory compactifications so
that our results will apply to both in the large moduli limit.

\subsection{Soft terms}

For definiteness 
 we will assume here that the SM resides at stacks of 7-branes wrapping
locally small cycles (of size $t$) in a CY whose overall volume is controlled by 
a large modulus $t_b$. We can  model out this structure with
a Kahler potential of the form \cite{conlon,ccq}
\beq
G\ =\ -2\log(t_b^{3/2} \ -\ t^{3/2}) \ +\ \log|W|^2\, ,
\eeq
with $t=T+T^*$ being the relevant local modulus. 
The gauge kinetic function and Kahler metrics of a MSSM matter field $C_\alpha$ 
will be given  to leading order by
\footnote{We will ignore here the dependence of the Kahler
potential on the axidilaton and complex structure fields
 since they play no role in the computation of the soft terms
under consideration, at least to leading order in $\alpha '$.}

\beq
f_a\ =\ T \ \ \ ;\ \ \  K_\alpha \ =\ \frac {t^{(1-\xi_\alpha )}
}{t_b}\, . 
\eeq
It is important to emphasize that what is relevant here is the no-scale
structure of the moduli Kahler potential and that analogous results would 
be obtained in a more general CY or F-theory compactification 
 with more  Kahler moduli as long as we assume
that our local modulus $t$ is is
smaller than the overall modulus $t_b$ so that an expansion on $t/t_b$ is consistent. 
 As emphasized in \cite{conlon,ccq} the
soft terms obtained for the local MSSM 7-branes will only depend on the local modulus $t$
and the value of its auxiliary field $F_t$ and will not directly depend on 
the large moduli. 
Note that, being a Kahler modulus, assuming $F_t\not=0$   corresponds to the assumption 
of the presence of non-vanishing SUSY-violating antisymmetric ISD fluxes \cite{gkp}
in the region in compact space where the SM 7-branes reside.

The MSSM superpotential $W$ has the general form
\beq
W\ =\ W_0 \ +\ \frac {1}{2} \mu H_uH_d \ +\ \frac {1}{6}
Y^{(0)}_{\alpha \beta \gamma } C^\alpha C^\beta C^\gamma \ +\ \ldots
\label{super}
\eeq
Here $W_0$ is a moduli dependent superpotential (including 
typically a flux-induced constant term which controls the scale of
SUSY breaking). The $C^\alpha $ are  the 
SM chiral superfields and $H_{u,d}$ the minimal Higgs multiplets.
In our computations below we are going to assume that there is an explicit
$\mu$-term in the Lagrangian which can be taken (at least in some approximation)
as independent of the Kahler moduli $t_b,\ t$.
A simple origin for such a term  is again RR and NS fluxes.
Indeed, it is known that certain SUSY preserving fluxes may give rise to  explicit  
supersymmetric
mass terms to chiral fields, $\mu$-terms
(see e.g. \cite{grana,ciu1,ciu2,fluxed}). If in addition there are SUSY-breaking
fluxes a $B$-term and the rest of soft terms appear. These two kind of fluxes
are  in principle  independent so that one can simply consider an explicit
constant (moduli independent) $\mu$-term in the effective action as a free parameter.
On the other hand, if both the origin of the $\mu$-term and SUSY-breaking
soft terms is fluxes,  it is reasonable to expect both to be of the same
order of magnitude, solving in this way the $\mu$ problem
 \footnote{An alternative
to get a Higgs bilinear could be the presence of an
additional SM singlet chiral field $X$  coupling to the Higgs fields
like $XH_{u}H_d$. It is trivial to generalize all the above formulae to this
NMSSM case.}.
However,  
 in the general formulae below
we will  allow for  the presence of a
Giudice-Masiero term  of the form $Z(T,T)H_uH_d+ h.c.$ in the Kahler
potential \cite{gm}. A discussion on alternative  origins  for
a $\mu$-term in string models is also given in the appendix. 
As we describe there, these alternatives do not look particularly 
promising concerning correct radiative EW symmetry breaking and hence
in  the
phenomenological analysis below we will  set $Z=0$.

Now the form of the effective soft Lagrangian obtained 
 is given by (see e.g.\cite{bim})
\begin{eqnarray}
{\cal L}_{soft} &=& \frac{1}{2}(M_a 
{\widehat{\lambda}}^a {\widehat{\lambda}}^a + h.c.)
- m_{\alpha}^2 \widehat{C}^{*\overline {\alpha}}
\widehat{C}^{\alpha}
\nonumber\\ &&
-\
\left(\frac{1}{6} A_{\alpha \beta \gamma} \widehat{Y}_{\alpha \beta \gamma}
	    \widehat{C}^{\alpha} \widehat{C}^{\beta} \widehat{C}^{\gamma}
  + B {\widehat{\mu}} \widehat{H}_u \widehat{H}_d+h.c.\right)\ ,
\label{F6}
\end{eqnarray}
with
\begin{eqnarray}
M_a &=&\frac{1}{2}\left(Ref_a\right)^{-1} F^m \partial_m f_a  \; ,
\label{F}\\
{m}_{\alpha}^2 &=& 
\left(m_{3/2}^2+V_0\right) - {\overline{F}}^{\overline{m}} F^n 
\partial_{\overline{m}}\partial_n \log{ K_{\alpha}}\ ,
\label{mmmatrix}
\\
A_{\alpha\beta\gamma} &=& 
F^m \left[  { K}_m + \partial_m \log Y^{(0)}_{\alpha\beta\gamma} 
- \partial_m \log({ K_{\alpha}} { K_{\beta}}
{  K_{\gamma}}) \right]\ ,
\label{mmmatrix2}
\\
B &=& {\widehat{\mu}}^{-1}({ K}_{H_u}{\ K}_{H_d})^{-1/2}
\left\{ \frac{ { W}^*}{|{ W}|} e^{{ K}/2} \mu 
\left( F^m \left[ {\hat K}_m + \partial_m \log\mu\right.\right.\right.
\nonumber\\ && 
\left.\left.-\ \partial_m \log({ K_{H_u}}{ K_{H_d}})\right]
- m_{3/2} \right)
\nonumber\\ &&
\ + 
\left( 2m_{3/2}^2+V_0 \right) {Z} -
m_{3/2} {\overline{F}}^{\overline{m}} \partial_{\overline{m}} Z
\nonumber\\ &&
+\ m_{3/2} F^m \left[ \partial_m Z - 
Z \partial_m \log({ K_{H_u}}{ K_{H_d}})\right]
\nonumber\\ &&
\left.-\ {\overline{F}}^{\overline{m}} F^n 
\left[ \partial_{\overline{m}} \partial_n Z - 
 (\partial_{\overline{m}} Z) 
\partial_n \log({ K_{H_u}}{ K_{H_d}})
\right] \right\}
\ ,  
\label{mmmatrix3}
\end{eqnarray}
where $V_0$ is the vacuum energy which is assumed to be negligibly small
from now on.
Here $\widehat{C}^{\alpha}$ and $\widehat{\lambda}^a$ are the scalar and
gaugino canonically
{\it normalized} fields respectively
\begin{eqnarray}
\widehat{C}^\alpha &=& { K}_{\alpha}^{1/2} C^\alpha\ ,
\label{yoquese}
\\
\widehat{\lambda}^a &=& (Re f_a)^{1/2} \lambda^a\ , 
\label{F7}
\end{eqnarray}
and the rescaled Yukawa couplings and $\mu$ parameter 
\begin{eqnarray}
{\widehat{Y}}_{\alpha \beta \gamma}    &=& Y_{\alpha \beta \gamma}^{(0)} \; 
 \frac{ { W}^*}{|{ W}|} \; e^{{ K}/2} \;
({ K}_\alpha { K}_\beta { K}_\gamma)^{-1/2}\ ,
\label{yyoquese}
\\
\widehat{\mu} &=& 
\left(  \frac{ { W}^*}{|{ W}|} e^{ K/2} {\mu}
+ m_{3/2} Z -
{\overline {F}}^{\overline{m}} \partial_{\overline{m}} Z \right)
({ K}_{H_u}{ K}_{H_d})^{-1/2}\ ,
\label{rescalado}
\end{eqnarray}
have been factored out in the $A$ and $B$ terms as usual. 

In Type IIB orientifolds the holomorphic 
perturbative superpotential is independent of the
Kahler moduli  so that
the derivatives of ${\widehat Y}^{(0)}$  
 in the expression for $A$   vanish.
Using these  formulae one then obtains general  soft terms 
(for $t_b\gg t$ ) as follows:
\beqa
M\ & = & \frac {F_{t}}{t}\, , \\
m_\alpha ^2\ & = &\ (1-\xi_\alpha ) |M|^2
 \ \ , \ \  \alpha =Q,U,D,L,E,H_u,H_d \, ,\ \\ \nonumber
A_U\  & = & \ -M(3-\xi_{H_u}-\xi_Q-\xi_U)\, , \\ \nonumber
A_D\  & = & \ -M(3-\xi_{H_d}-\xi_Q-\xi_D)\, , \\ \nonumber
A_L\  & = & \ -M(3-\xi_{H_d}-\xi_L-\xi_E)\, , \\ \nonumber
B \   & = & \ -M(2-\xi_{H_u}-\xi_{H_d}) \ .
\label{cojosoftgen2}
\eeqa
This is analogous to results found in \cite{fluxed} for the 
case of a single overall modulus $T$.
Note that the dependence on the SUSY-breaking from the large
modulus $t_b$ disappears to leading order and the size of soft terms
is rather controlled by the local  modulus $t$. In particular, gaugino masses
corresponding to the SM gauge groups depend only on the modulus of
the local 4-cycles they wrap, rather than the large volume modulus $t_b$.
These gaugino masses set the scale of the SUSY-breaking soft terms.
As in \cite{conlon,ccq} here the gravitino mass will be given approximately 
by the auxiliary field of large moduli, $m_{3/2}\simeq -F_{t_b}/t_b$,
with corrections suppressed by the large volume $t_b$.
One can check that all  the  bosonic soft terms above 
may be understood as
coming from the positive definite scalar potential \cite{fluxed}
\beq
\ V_{SB}\ =\ \sum_\alpha  \ (1-\xi_{\alpha}) \vert \partial_\alpha  W \ -\ M^*C_{\alpha }^*\vert ^2
\ +\ \sum_\alpha  \xi_\alpha  \vert \partial_\alpha  W \vert ^2\, .
\label{complimas}
\eeq
The positive definite structure may be seen as a consequence of the
'no-scale' structure of modulus domination and is expected
 to apply also in more general  F-theory compactifications.

Within the philosophy of gauge coupling unification we will 
assume unified modular weights:
\beq
\xi_f \ =\ \xi_Q =\xi_U =  \xi_D =\xi_L =\xi_E\, .
\eeq
This is also reasonable within e.g. an F-theory approach 
with an underlying GUT-like symmetry like $SO(10)$ in
 which one expects all fermions to have the same  modular weight.
In us much as the effect of magnetic fluxes giving rise to
chirality for SM fermions is negligible (see next section) one also expects
flavour independence for them. So we will assume a universal
modular weight $\xi_f$ for all quarks and leptons.
Concerning the Higgs multiplets we have seen that they 
can have $\xi_H=0,1,1/2$ and hence there is no reason why 
they should have the same modular weight as chiral fermions.
We will however assume that  both Higgses have the same
modular weight $\xi_H=\xi_{H_u}=\xi_{H_d}$, as would also be expected in
models with an underlying left-right symmetry.

Under these conditions the summary of the soft terms 
obtained for 
the three possibilities for brane distributions with
consistent Yukawa couplings, (A-A-$\phi$) , (I-I-A) and (I-I-I)
are shown in  table 1.

\begin{table}[htb] \footnotesize
\renewcommand{\arraystretch}{1.50}
\begin{center}
\begin{tabular}{|c|c||c|c|c|c|c|c|}
\hline  $(\xi_L,\xi_R,\xi_H)$   &  Coupling &
   M  &  $m_L^2$
 &  $m_R^2$    &  $m_H^2$
 &   A     &  $B$     \\
\hline\hline
 $(1,1,0)$ &   (A-A-$\phi$) &
$M$   &   0  &    0     &   $ {|M|^2}$   &  $-M$  &  $-2M$
   \\
\hline
 $(1/2,1/2,1)$ &   (I-I-A)  &
$M$   &   $\frac {|M|^2}{2}$    &  $\frac {|M|^2}{2}$   
  &   0    &  $-M$  &  0    \\
\hline
$(1/2,1/2,1/2)$ &   (I-I-I)  &
$M$   &   $\frac {|M|^2}{2} $  
& $\frac {|M|^2}{2} $ &  $\frac {|M|^2}{2}$    &  -3/2M    &  $-M$      
\\
\hline \end{tabular}
\end{center} \caption{\small Modulus   dominated soft terms
for  choices of modular weights $\xi_\alpha$  which are consistent with
the existence of trilinear Yukawa couplings in $7$-brane systems. }
\label{opciones reales}
\end{table}

Note that in the scenarios with couplings (A-A-$\phi$) and (I-I-A) it is natural to
assume that the Higgs field is identified with fields of type 
$\phi$ and $A$ respectively and these are the cases displayed in the table.
Concerning the $B$ parameter it is obtained assuming an explicit 
$\mu$-term.

We will study the phenomenological viability of these
three options in the next chapter. Since the cases with couplings
(I-I-A) and (I-I-I) only differ in the origin of the Higgs fields, and
hence in the values of their modular weights, they will
give rise to a similar phenomenology. For this reason, we will analyse
the general case in which the Higgs modular weight is a free parameter,
$\xi_H$, and regard examples
(I-I-A) and (I-I-I) as the limiting cases with $\xi_H=1$ and
$\xi_H=1/2$, respectively.
Notice that this can be
understood as if the physical MSSM Higgs was a linear
combination of two fields with the correct quantum numbers, one of
them living in the intersection of two $D7$-branes and the other one in
the bulk of one of them 
\footnote{ In chapter 6 in ref.\cite{bhv} an F-theory  $SO(10)$ model
is constructed in which indeed there are massless Higgs multiplets 
both in the bulk and at intersections.  }

On the other hand, the case with couplings (A-A-$\phi$) is unrelated
to the previous ones and will be studied separately.

\subsection{ Corrections to soft terms from magnetic fluxes}

The Kahler metrics and gauge kinetic functions used  in the computations above 
correspond to the leading behaviour in $\alpha '$. It is  interesting to
estimate what could be the effect of subleading terms  coming from 
possible magnetic fluxes in the 7-branes, as discussed in subsection  2.5.
We know that the presence of such fluxes is required to get chirality.
To do that we can use the results for the Kahler metrics given in the
diluted flux approximation in eq.(\ref{flujillas}). One finds for the 
soft terms the results:
\beqa
M \ &=&\ \frac{F_t}{t+as}\, ,\\
m_\alpha ^2\ & = & \  \frac {|F_t|^2}{t^2}  (1-\xi_\alpha)\ 
\frac { (1+ \frac {c_{\alpha}\xi_\alpha }{t^{(1-\xi_\alpha )}})}
{(1+ \frac {c_\alpha }{t^{(1-\xi_\alpha )}})^2}\, ,
\\
A_{\alpha \beta \gamma}\ & = & \ -\ \frac {F_t}{t} \ \sum_{i=\alpha, \beta, \gamma}
\ (1-\xi_i)(1-\frac {c_i}{t^{(1-\xi_i)}})\, ,\\
B\ &=&\ -\frac {F_t}{t} \sum_{i=H_u, H_d} \ (1-\xi_i)(1- \frac {c_i}{t^{1-\xi_i}} ) \, .
\eeqa
Note that for matter fields coming from a $D=8$ vector multiplet (modular weight $\xi=1$) 
the scalar terms are still zero and get no flux correction.
In the computation of the $B$-term an explicit $\mu$ term independent on $t,t_b$ 
has been assumed. 
We will make use of these formulae to try to estimate the
effect of fluxes on the obtained low energy physics below.

\subsection{Soft terms and moduli fixing}

In the above computations we have tacitly assumed that all moduli have been fixed 
at  a (slightly) de Sitter vacuum of the scalar potential. The issue of 
moduli fixing in Type IIB orientifold compactifications has been addressed 
by many authors in recent years starting with the work in \cite{kklt}
(see \cite{reviewsfluxes} for reviews and references).
Concerning the computation of soft terms in a model with 
moduli stabilized, two situations have received 
special attention, 'Mirage mediation' and 
'Large Volume Compactification' (LVC) models.
 In the case \cite{kklt} of a KKLT model with a
single overall modulus, it has been argued \cite{mirage} that the size of the
modulus dominance contribution to SUSY-breaking  $m_{soft}$ is
suppressed relative to the gravitino mass as
$m_{soft}\simeq  m_{3/2}/\log(M_p/m_{3/2})$.
Then one cannot neglect the anomaly mediation contribution 
to SUSY-breaking which is competitive with modulus
dominance. The  hierarchy $M_p/m_{3/2}$ comes 
from an assumed  small value for the (flux-induced) 
  superpotential $W_0$.
This situation goes under the name of
'mirage mediation'. 
In this situation one would have to add
to the soft terms here computed  the 
well known universal contribution from 
anomaly mediation.

In the LVC scenario 
the prototype of CY considered in these large volume 
models \cite{conlon,acqs} 
has a Kahler potential like that in
eq.(\ref{dosmoduli}).
In this scheme the hierarchy $M_p/m_{3/2}$ is large due
to the presence of the large volume cycle controlled 
by $t_b$ and the string scale is at
an intermediate value, $M_s\propto 10^{11}$ GeV.
 Such a large value for $t_b$ is obtained as
a result of a minimization of a scalar potential 
in which a one loop term in the $\alpha '$ expansion
is also included.  The flux induced 
superpotential $W_0$ is of order one and 
the large value for $t_b$ is due to a combination 
of non-perturbative instanton 
corrections  and one-loop terms in the $\alpha '$ expansion.
 In this scheme SUSY-breaking is
dominated by a modulus dominance pattern as discussed 
in the previous sections.

As discussed in \cite{acqs}, as one decreases $|W_0|$ in the
LVC scheme one ends up in a situation rather analogous to
mirage mediation. There are however other  small $|W_0|$
minima  which are  compatible with a large value of the 
string scale (of order of the canonical GUT scale, $10^{16}$ GeV)  
and modulus dominated SUSY-breaking. Such a large string scale 
scenario  would be interesting if one wants to understand the
joining of coupling constants in a simple way.
In the numerical analysis in the next section we will be
tacitly
assuming  a situation similar to that, 
with modulus dominance and 
a string scale of order the GUT scale.
In particular, in this case the anomaly mediation contribution is
suppressed by the same no-scale cancellation which makes the dependence
on the large modulus $t_b$ to disappear from the final expression for soft terms 
\cite{conlon,acqs,ccq},

Before presenting our study of the viability of the
three different options for modulus  domination that were introduced 
in the previous sections
let us make  
a few comments  concerning some recent computations of soft terms
in string models. In \cite{fluxed} the cases with $(\xi_H,\xi_L,\xi_R)=(0,0,0)$
and $(1/2,1/2,1/2)$ were discussed. The phenomenological viability of the first
was analyzed in detail in \cite{abi}. However, there is no known perturbative
compactification with 
such modular weight distribution which is 
compatible with the existence of trilinear couplings.
Concerning the $(1/2,1/2,1/2)$ choice, no phenomenological analysis
has been yet reported \footnote{An intersecting brane model with a 'dilute fluxed
limit' corresponding to those modular weights was analyzed in \cite{fi}
but no phenomenological analysis was made. See however
\cite{Gordy}.}.
As we said, 
in \cite{mirage} it was found  that in 
one-modulus 
KKLT-like vacua one is lead to a
mixed situation with soft terms coming from competing contributions  from
modulus dominance and anomaly mediation. Phenomenological analysis have been made
(see e.g. \cite{tata,choi}). 
It would be interesting to reconsider those analysis
by including the different allowed options for modular weights.
Finally, 
in \cite{conlon} soft terms have been computed in a LVC
2-modulus scenario with an intermediate string scale.
In this case one has modulus domination and our results would 
apply. The anomaly mediation contribution is negligible.
The authors however concentrate on universal  
values of the modular weights $\xi =2/3$ and again no
 example of compactification with such modular weights is known. 
Another difference with our analysis is that in that case the
string scale is at an intermediate scale and 
the soft terms are then run from an intermediate scale down to the weak scale.
In our analysis we   will assume
a large string scale of order the GUT scale.
For a recent phenomenological study of LHC signatures for soft terms
coming from different options for moduli fixing see also \cite{kks}.

\section{EW symmetry breaking and SUSY spectrum}

We are now ready to extract the low energy implications of the 
MSSM soft terms listed in table 1.
We will use those values as boundary conditions at the 
string scale which we will identify with the standard GUT scale 
at which the MSSM gauge couplings unify. We will solve  
numerically the  Renormalization Group Equations (RGEs) 
and calculate the low energy SUSY spectrum. We will also
impose standard radiative electroweak symmetry breaking.

\subsection{Radiative EW symmetry breaking and experimental constraints}

The minimization of the loop-corrected Higgs potential leaves the
following two conditions, which are imposed at the SUSY scale, 
\begin{eqnarray}
  \mu^2 & = & 
  \frac{ - m_{H_u}^2\tan^2\beta + m_{H_d}^2}{\tan^2\beta-1} 
  - \frac{1}{2} M_Z^2 ,
  \label{muterm}\\
  \mu B & = &  \frac{1}{2} \sin 2 \beta \, (m_{H_d}^2 +m_{H_u}^2 + 2
  \mu^2)\,, 
  \label{Bterm}
\end{eqnarray}
where $\tan\beta\equiv\langle H_u \rangle/\langle H_d \rangle$ is the
ratio of the Higgs vacuum expectation values and $m_{H_{u,d}}^2$
correspond to the Higgs mass parameters shifted through tadpole
terms. 
It should be noted that our choice for the sign of the $\mu$
parameter, consistent with our convention for the superpotential
(\ref{super}), is opposite to the usual one.

A usual procedure consists then in fixing the value of $\tgb$ and
use the experimental result for $M_Z$ to determine the modulus of the
$\mu$ parameter via eq.(\ref{muterm}). The $B$ parameter is then
obtained by solving 
eq.(\ref{Bterm}).
In this approach, once the modular weights which describe the specific
model are given, 
the only free parameters left are the common gaugino
mass, $M$, the value of $\tan\beta$ and the sign of the $\mu$
parameter (not fixed by condition (\ref{muterm})).

On the other hand, 
given that the value of the bilinear soft term, $B(M_{GUT})$, is also
a prediction in these constructions (dependent on the source of the
$\mu$ term) a more complete approach consists in 
imposing it as a boundary condition for the RGEs at
the string scale. Conditions (\ref{muterm}) and (\ref{Bterm}) can
then be used to determine
both $\tgb$ and $\mu$ as a function of the only free parameter, $M$. 
Note that this is a extremely constrained situation and it is not
at all obvious a priori that solutions passing all experimental
constraints exist.

It is not possible to derive an analytical solution for
$\tan\beta$ from eqs.(\ref{muterm}) and (\ref{Bterm}), 
since tadpole corrections to the Higgs mass terms depend on
$\tan\beta$ in a non-linear way. Furthermore, $\tan\beta$ is needed in
order to adjust the Yukawa couplings at the GUT scale so that they
agree with data.
Thus, to find a solution an iterative procedure has to be used where 
the RGEs are numerically solved
for a first guess of $\tan\beta$ using the soft
parameters as boundary conditions at the GUT scale. 
The resulting $B$ at the
SUSY scale is then compared with the solution of eq.(\ref{Bterm}).
In subsequent iterations, the value of $\tan\beta$ is varied, 
looking for convergence of the resulting $B(M_{SUSY})$. 
It is not always possible to find a solution with consistent REWSB, 
and this excludes large areas of the parameter space \footnote{
  Moreover, the same input values for the soft parameters can lead to
  multiple solutions in terms of $\tan\beta$ \cite{dn91}. We have
  explicitly checked this is not the situation in any of our examples
  for $\mu<0$, but usually occurs for $\mu>0$.}. 
In our analysis we have implemented such iterative process through a
modification of the {\tt SPheno2.2.3} code \cite{spheno}. 
The constraints imposed by the full REWSB conditions 
have been analysed for models with 
universal soft parameters \cite{emy02,eoss}, as
well as string-inspired scenarios \cite{abi}.

Once the supersymmetric spectrum is calculated, compatibility with
various experimental bounds has to be imposed. 
We have taken into account the constraints obtained by LEP on 
the masses of supersymmetric particles, as well as on the lightest
Higgs boson \cite{pdg}.
Also, the most recent experimental limits on the 
contributions to low-energy observables have been included
in our analysis.
More specifically, we impose the experimental
bound on the branching ratio of the rare $\bsg$ decay,
$2.85\times10^{-4}\le\,{\rm BR}(\bsg)\le 4.25\times10^{-4}$, obtained
from the experimental world average 
reported by the Heavy Flavour Averaging Group \cite{bsgHFAG07},
and the theoretical calculation in the Standard Model
\cite{bsg-misiak}, 
with errors combined in quadrature.  
We also take into account the upper constraint on
the $(\bmumu)$ branching ratio obtained by CDF,
BR$(\bmumu)<5.8\times10^{-8}$ at $95\%$ c.l. \cite{bmumuCDF07}
(which improves the previous one from D0 \cite{bmumuD007}).

Regarding the muon anomalous magnetic moment, a constraint on the
supersymmetric contribution to this observable, $\asusy$, can be
extracted by comparing the experimental result
\cite{g-2}, with the most recent theoretical evaluations of the
Standard Model contributions \cite{g-2_SM,newg2,kino}. 
When $e^+e^-$ data are used the experimental excess in
$a_\mu\equiv(g_\mu-2)/2$ would constrain a possible supersymmetric
contribution to be $\asusy=(27.6\,\pm\,8)\times10^{-10}$, where
theoretical and experimental errors have been combined in
quadrature. However, when tau data are used, a smaller discrepancy
with the experimental measurement is found. Due to this reason, in our
analysis we will not impose this constraint, but only indicate the
regions compatible with it at the $2\sigma$ level, this is, 
$11.6\times10^{-10}\le\asusy\le43.6\times10^{-10}$.

Assuming R-parity conservation, and hence the stability of the LSP, we
also investigate the possibility of obtaining viable neutralino dark
matter. This is, in the regions of the parameter space where the
neutralino is the LSP we compute its relic density by means of the
program {\tt micrOMEGAs} \cite{micromegas}, and check compatibility
with 
the data obtained by the
WMAP satellite \cite{wmap5yr}, which constrain the amount of cold
dark matter to be 
$0.1037\le\Omega h^2\le 0.1161$.

The value of the mass of the top quark is particularly relevant. 
In our computation we have used the central
value corresponding to the
recent measurement by CDF \cite{topCDF08},
$m_t=172 \pm 1.4\,{\rm GeV}$. 
We will briefly comment on the effect that 
deviations from this quantity may have on REWSB.

Finally, the presence in SUSY theories of scalar fields which
carry electric and colour charges can lead to the occurrence of
minima of the Higgs potential where charge and/or colour 
symmetries are broken when
these scalars take non-vanishing VEVs. If
these minima are deeper than the physical (Fermi) vacuum, the latter
would be unstable. The different directions in the field space that
can lead to this situation were analysed and classified in
\cite{clm1}. It was found there that the most dangerous direction
corresponds to the one
labelled as UFB-3, where the stau and sneutrino take non-vanishing
VEVs, since the tree-level scalar potential could even become
unbounded from below. 
These UFB constraints were found to impose stringent constraints on
the parameter space of general supergravity theories \cite{cggm03-1},
as well as in different superstring
and M-theory scenarios \cite{ibarra,ckm08}.
In our study we will comment on the constraints which are
derived when the absence of such charge and/or colour-breaking minima
is imposed \footnote{
  Strictly speaking, 
  the existence of a global charge and/or colour-breaking vacuum cannot
  be excluded if the lifetime of the metastable physical minimum is
  longer than the age of the Universe, as is usually the
  case. 
}.

In order to understand the effect of all these experimental and
astrophysical constraints we have performed a scan over the gaugino
mass parameter, $M$, and $\tan\beta$ for the three different
consistent models that were specified in Table\,\ref{opciones
  reales}. 
We will discuss first the results obtained without imposing the
predicted boundary condition for $B$ and analyse the effect of this
constraint 
in the following subsection.

\subsection{ The intersecting 7-brane (I-I-I)-(I-I-A) configurations}

As commented at the end of Section\,3.1, we will analyse  these two cases
together within the framework of a generic scenario
in which the modular weight for the Higgs is a free parameter. 
In this  approach, $\xi_H$ can take any value
between $\xi_H=1/2$, which corresponds to case (I-I-I), and $\xi_H=1$,
as in case (I-I-A). The soft parameters for such a model can be
extracted from (\ref{cojosoftgen2}) and read
\begin{eqnarray}
m_{L,E,Q,U,D} ^2\ & = &\ |M|^2/2\, ,\ \\ \nonumber
m_{H_u,H_d} ^2\ & = &\ (1-\xi_H )|M|^2\, ,\ \\ \nonumber
A_{U,D,L}\  & = & \ -M(2-\xi_{H})\, , \\ \nonumber
B \   & = & \ -2M(1-\xi_{H}) \, .
\label{soft-iih}
\end{eqnarray}

\begin{figure}[t!]
  \epsfig{file=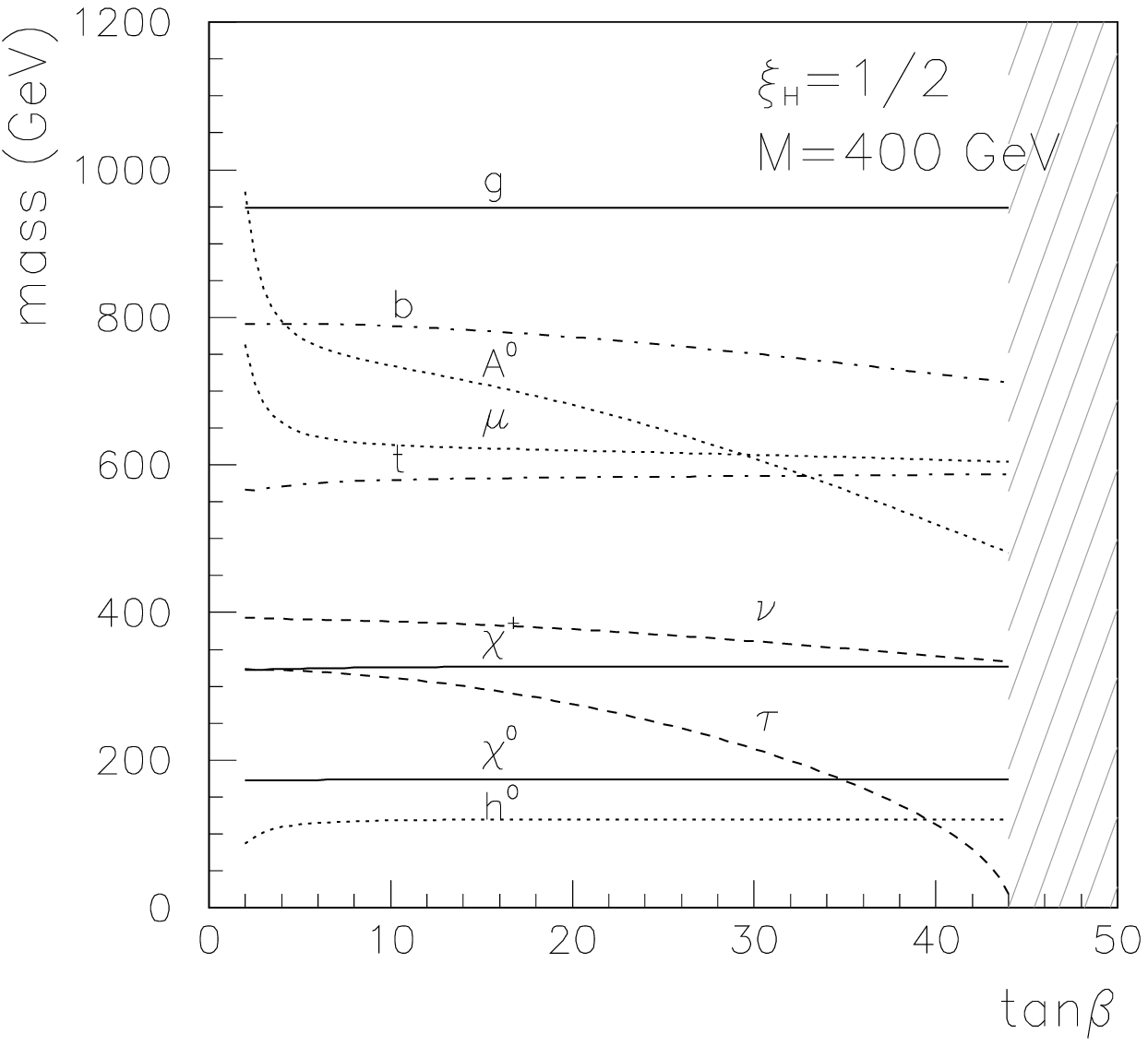,width=8cm}
  \epsfig{file=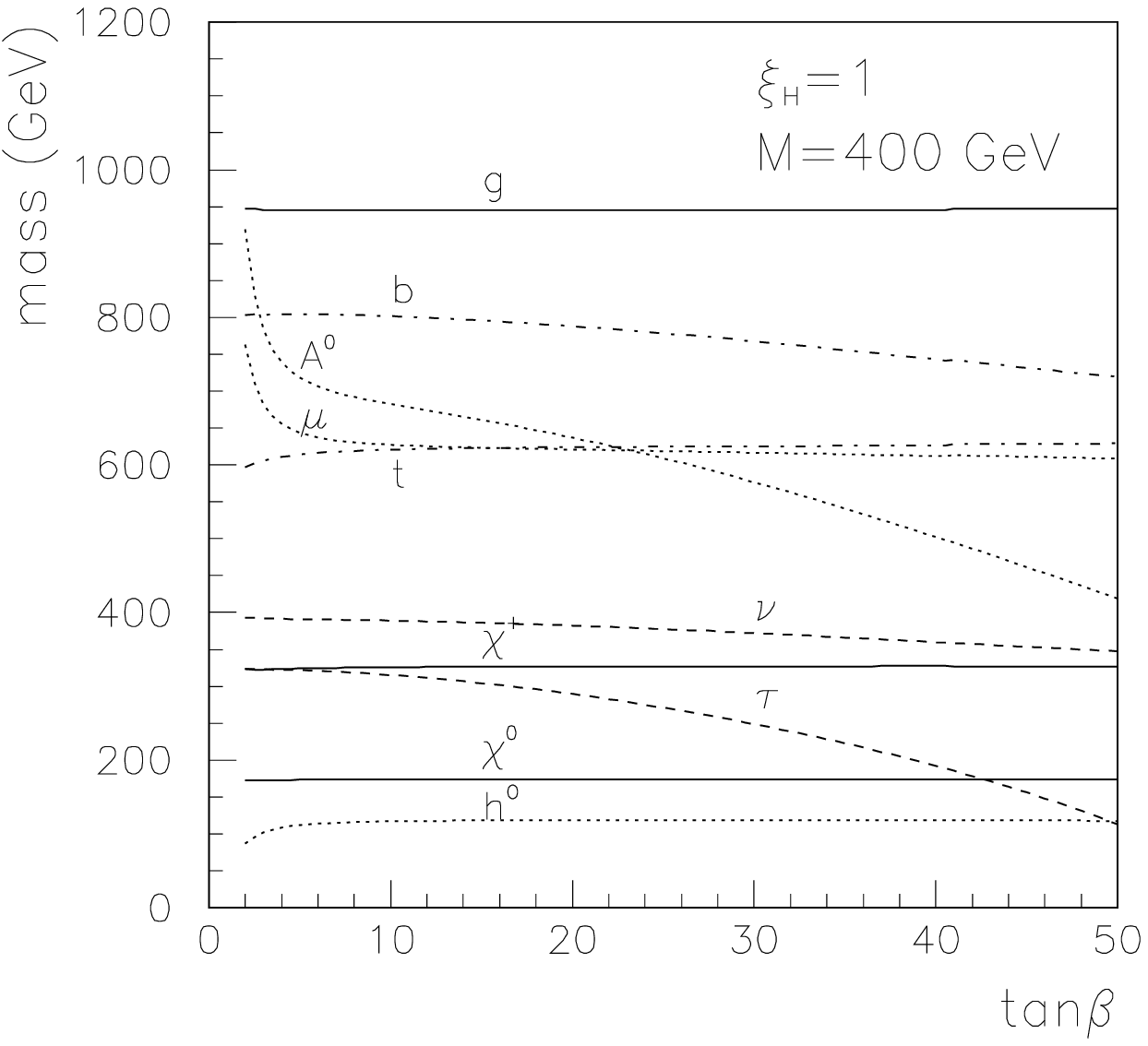,width=8cm}
  \vspace*{-1cm}
  \caption{
    Low-energy supersymmetric spectrum as a function of $\tan\beta$
    for $\xi_H=1/2,$ (left) and $\xi_H=1$ (right) with 
    $M=400\ {\rm GeV}$ and $\mu<0$. From bottom to top, the solid lines
    represent the masses of the lightest neutralino, the lightest
    chargino, and the gluino. Dashed lines display the masses of the
    lightest stau and lightest sneutrino. Dot-dashed lines correspond to
    the stop and sbottom masses. Finally, the lightest Higgs mass, the 
    pseudoscalar Higgs mass
    and the absolute 
    value of the $\mu$ parameter are displayed by means of
    dotted lines. The ruled area for large $\tan\beta$ is excluded by
    the occurrence of tachyons in the slepton sector. 
    \label{spectrum-iih}
  }
\end{figure}

A sample of the resulting supersymmetric spectrum is represented in
fig.\,\ref{spectrum-iih} as a function of the value of $\tan\beta$ for
$M=400\,{\rm GeV}$ and $\mu<0$ for the two limiting cases $\xi_H=1/2$
and $\xi_H=1$. 
As evidenced by the plot, despite the fact that the slepton 
mass-squared terms 
are always positive at the GUT scale, the running 
down to the EW scale can lead to the occurrence of
tachyonic eigenstates for large  tan$\beta$. 
This is typically the case of the
lightest stau.
The negative contribution to the
RGEs governing the stau mass parameters increases with $\tan\beta$,
since it is proportional to the 
lepton
Yukawa, which varies as $1/\cos\beta$. 
As a consequence, the stau mass decreases towards large values of 
$\tan\beta$,
first becoming the LSP (which in the $\xi_H=1/2$ example occurs for
$\tan\beta\gsim 35$), then falling below its experimental
lower bound and eventually turning tachyonic. This sets an
upper constraint on the possible value of $\tan\beta$ (which obviously
increases for larger values of $M$). 
The effect of this constraint is more important when $\xi_H$ is small,
since the trilinear parameter $A_L$ is larger (more negative)
and enhances the negative contribution in the RGEs of the
stau mass parameters. 
Thus, the bound $\tan\beta\lsim45$ for $\xi=1/2$ is relaxed to
$\tan\beta\lsim55$ with $\xi=1$.

Another interesting feature is that
the resulting value for $|\mu|$, calculated from 
eq.(\ref{muterm}), turns out to be relatively large,
of order $1.5\,M$.  
Similarly, the pseudoscalar Higgs (as well as the
heavy neutral and charged Higgses) is also heavy, decreasing for large
values of $\tan\beta$. When the value of $\xi_H$ increases the
predicted pseudoscalar mass is slightly smaller.

The rest of the properties of the spectrum are less sensitive to
variations in the Higgs modular weight. 
For small values of $\tan\beta$ the lightest neutralino is typically
the LSP in this example. 
Since the value of the $\mu$ parameter 
is always large,
the lightest neutralino is mostly bino-like.
The universality of gaugino masses at the GUT scale and 
the large values of $|\mu|$ 
also lead to 
the well known low-energy relation among the masses of the 
lightest neutralino, the lightest chargino and the gluino,
$m_{\tilde\chi^0}:m_{\tilde\chi^+}:m_{\tilde g}\approx1:2:5.5$. 
The squark sector is rather heavy, another consequence of the
universality of the soft masses.

Notice finally that 
the lightest Higgs mass decreases towards small
$\tan\beta$ (through the decrease of its tree-level value). 
Hence, the LEP lower constraint on the Higgs mass
leads to a lower bound on
the phenomenologically viable 
values of $\tan\beta$. This bound also depends on the
value of $M$, as we will see, since $M$ sets the overall scale for the
soft parameters and therefore determines the size of loop 
corrections to $m_h$.

\begin{figure}[t!]
  \hspace*{-0.6cm}
  \epsfig{file=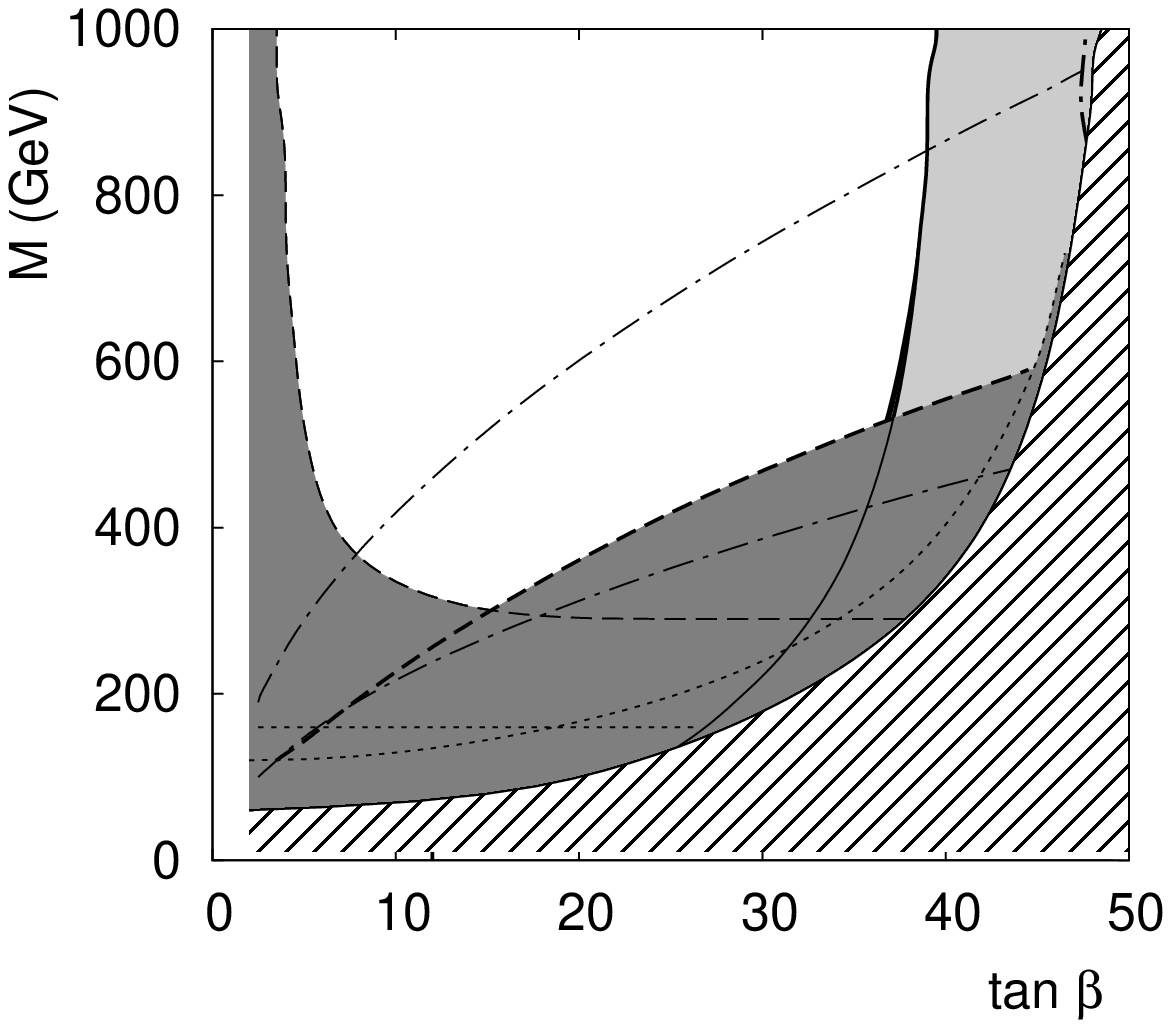,width=8.cm}
  \epsfig{file=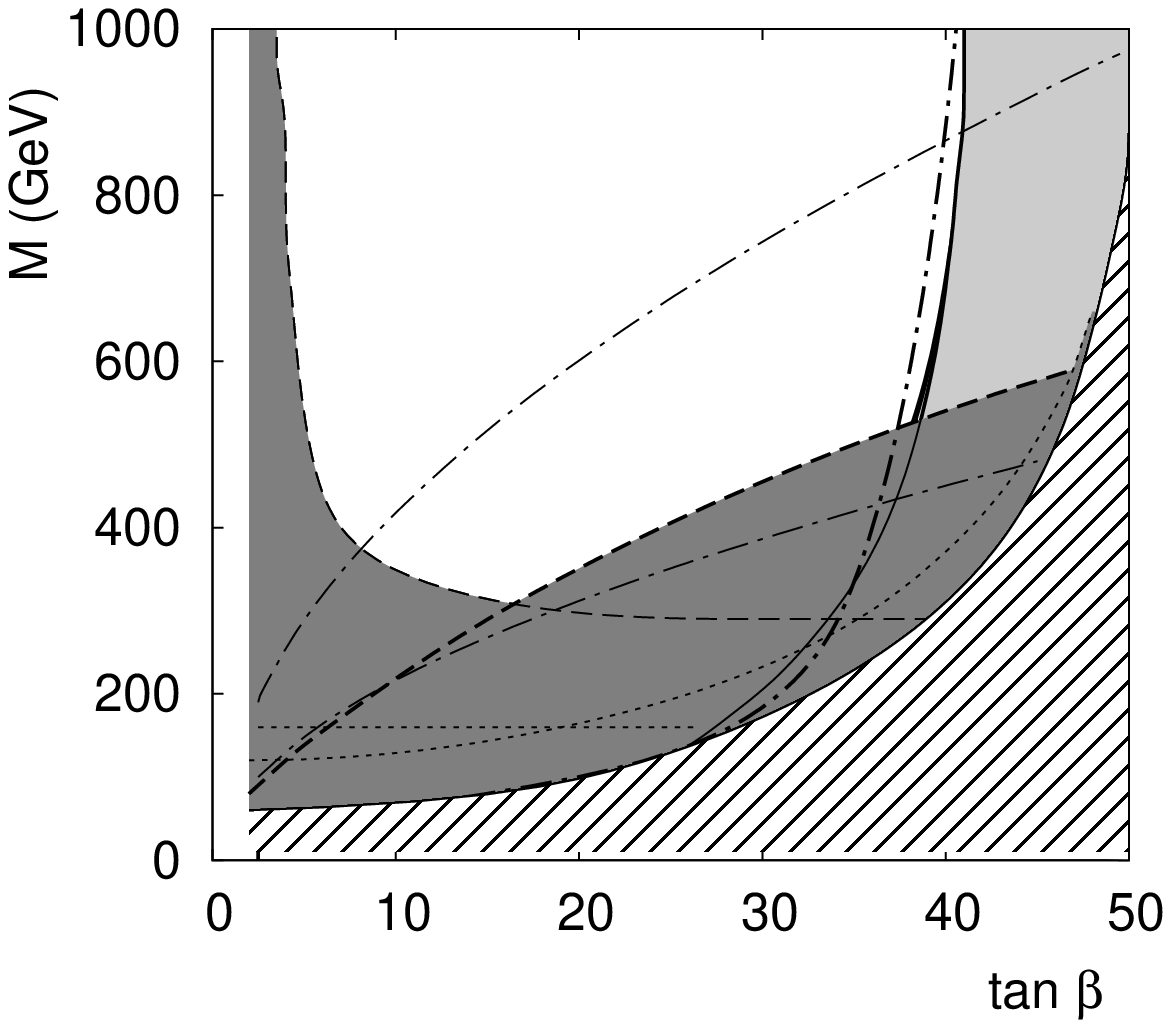,width=8.cm}
  \hspace*{-0.6cm}
  \epsfig{file=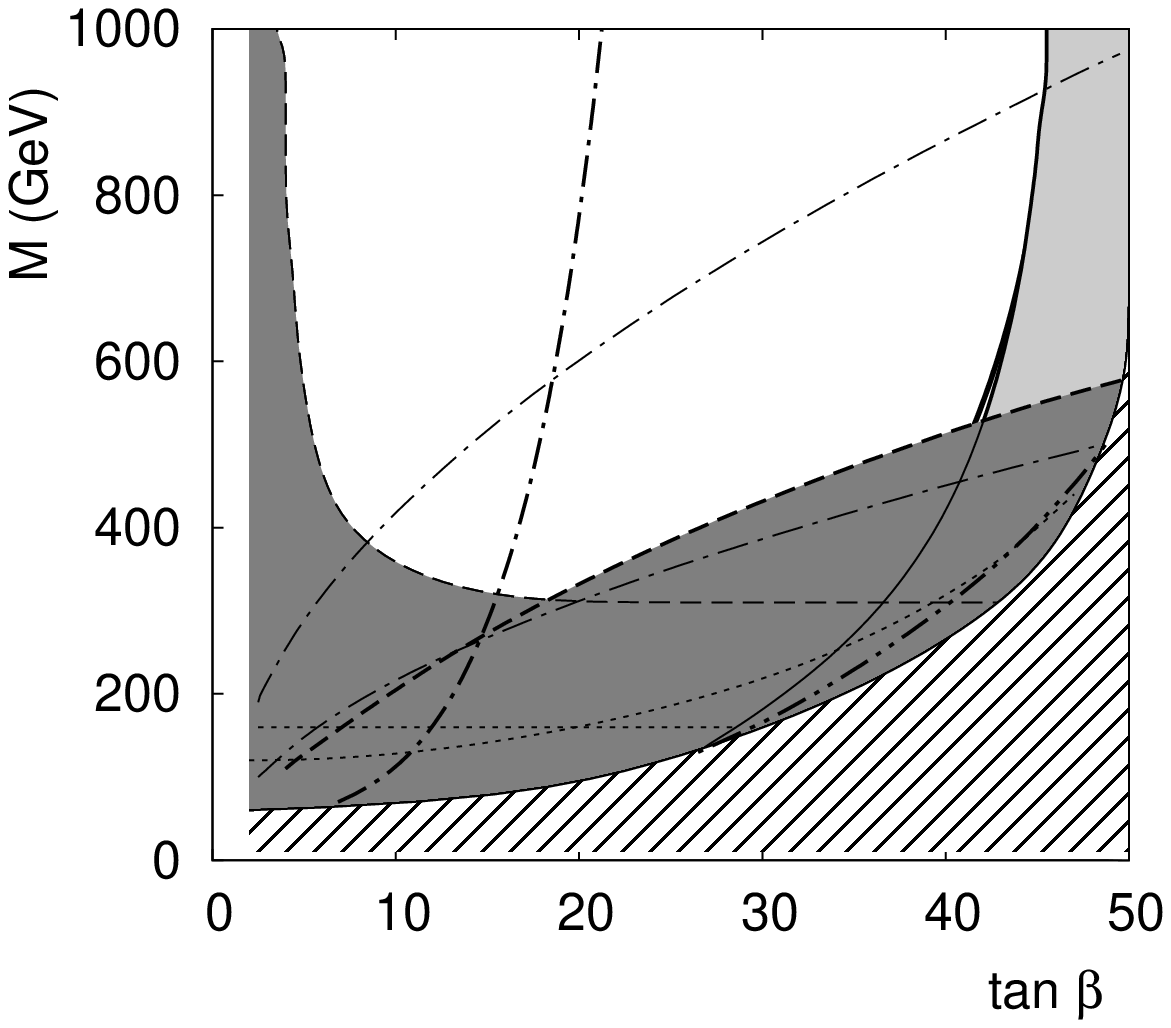,width=8.cm}
  \epsfig{file=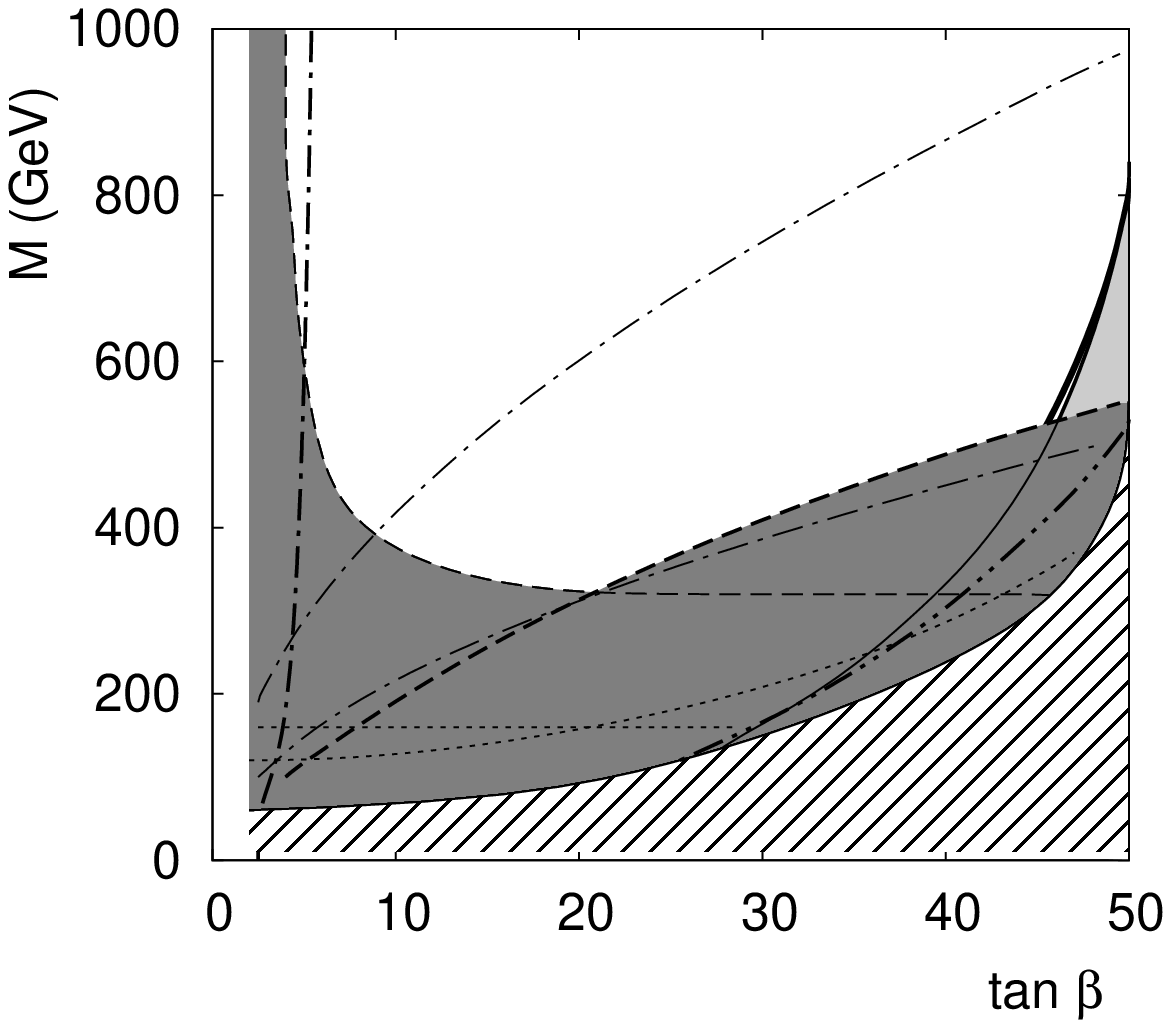,width=8.cm}
  \vspace*{-0.7cm}
  \caption{Effect of the various experimental constraints on the
    $(M,\,\tgb)$ plane for cases with $\xi_H=0.5,\,0.6,\,0.8$, and
    $1$, from left to right and top to bottom. Dark grey regions
    correspond 
    to those excluded by any experimental bound. 
    Namely, the area
    below (and to the left of) the thin dashed line is ruled out by
    the lower constraint on the lightest Higgs mass. 
    The region below
    the thin dotted lines is excluded by the lower bounds on the stau
    and chargino masses. 
    The area below the
    thick dashed line is excluded by $b\to s\gamma$.
    The region below the double dot-dashed line is excluded by
    $\bmumu$. 
    The thin dot-dashed lines correspond, from top to bottom, 
    to the lower and upper constraint on $\asusy$. 
    The area contained within solid lines corresponds to the region
    in which the stau is the LSP, and is depicted in light grey when
    experimental constraints are fulfilled. 
    In the remaining white area the neutralino is the LSP.
    The thin black area, in the vicinity of the region with stau LSP,
    corresponds to the region where the neutralino relic density is in
    agreement with the WMAP bound.  
    \label{mtgb-iih}
  }
\end{figure}

For a better understanding of the effect of the various experimental
constraints a full scan on the two free parameters $M$, and
$\tan\beta$ (for $\mu<0$) is presented in fig.\,\ref{mtgb-iih} for
four examples of Higgs modular weight \footnote{
  In these figures the constraint for the B-parameter has not yet
  been imposed. We will see below that the dark matter constrained 
  combined with the prediction for $B$ singles out the case with
  $\xi_H\simeq 0.5-06$.}, $\xi_H=0.5,\,0.6,\,0.8$ and $1$.
Some of the features
of the supersymmetric spectrum we have just described are also clearly
displayed.
For example, the ruled area for large $\tan\beta$ and small $M$
corresponds to the area excluded due to tachyons in the stau
eigenstates. This area becomes smaller as the modular weight for the
Higgs increases as a consequence of the increase in the stau mass.  
At the same time, the region with stau LSP (which is represented by
light grey) is also shifted towards
larger $\tan\beta$, thereby enlarging the area in which the neutralino
is the LSP.

The region below the thin dashed line is excluded by the LEP
constraint on the Higgs mass, and corresponds to $\tan\beta\lsim5$ and
$M\lsim300\,{\rm GeV}$. This constitutes the stronger lower limit
on $M$ for small values of $\tan\beta$. 
For $\tan\beta\gsim15$, however, 
the experimental bound on BR($\bsg$) sets a
more constraining lower limit on $M$, which increases with
$\tan\beta$ and reaches $M\gsim 500\,{\rm GeV}$.

The theoretical predictions for B($\bmumu$) only exceed the
experimental upper bound for small values of $M$ and very large 
$\tan\beta$. The reason is that, as already explained, the
pseudoscalar Higgs is always very heavy in these scenarios. 
Only in the cases with
$\xi_H=0.8$ and $\xi_H=1$ there are regions excluded for this reason. 
In any case, these areas are already disfavoured by a number of other
experimental bounds.

Regarding the supersymmetric
contribution to the muon anomalous magnetic moment, 
the region of the parameter space which is favoured by the
experimental constraint corresponds to the area between the thin
dot-dashed lines.  
The lower(upper) bound on $\asusy$ sets an upper(lower) limit on $M$. 
Since $\asusy$ increases for large
$\tan\beta$ (through an enhancement of the contribution mediated by
charginos and sneutrinos), 
both the upper and lower limits on $M$ also increase.
As we already indicated, in our analysis this constraint is not
imposed.

As a result, there are vast regions of the parameter space compatible
with all the experimental constraints and in which the lightest
neutralino is the LSP. 
In order to determine its viability as a dark matter candidate, its
relic density has to be 
computed and compared with existing bounds on the
abundance of cold dark matter.

As mentioned above, the neutralino is mostly bino in these
constructions, and for this reason 
its relic density easily exceeds the recent WMAP
constraint. The correct neutralino abundance is only found in those
regions of the parameter space where the neutralino mass is very close
to the stau mass, since then a coannihilation effect \cite{gs90} 
takes place
in
the early Universe which reduces very effectively the neutralino
abundance \footnote{Another possible mechanism 
that could help in reducing the neutralino relic
density is their resonant annihilation through an s-channel mediated
by a pseudoscalar Higgs. However, that is only possible when
$2m_{\tilde\chi^0}\approx m_{A^0}$. As we saw in 
fig.\,\ref{spectrum-iih} the pseudoscalar is too heavy in these
constructions for such a resonance to take place.}. 
After imposing the constraint on the neutralino relic density, the
only regions of the parameter space which are left correspond to very
narrow bands in the vicinity of the area with stau LSP. Interestingly,
these favour a very narrow range of values for $\tan\beta$, which is
always large. Also, as $\xi_H$ increases, the allowed region is
shifted towards larger $\tan\beta$. Thus, while
$35\lsim\tan\beta\lsim40$ for $\xi_H=1/2$ (case (I-I-I)), 
$45\lsim\tan\beta\lsim55$ is needed for  case (I-I-A) with
$\xi_H=1$.

So far we have not commented on the effect of UFB constraints. 
Most of the parameter space turns out to be disfavoured on these
grounds.
The reason for this is the low value of
the slepton masses, and more specifically, of the stau mass. Indeed,
the lighter the stau, the more negative the scalar potential along the
UFB-3 direction becomes, thus easily leading to a minimum deeper than
the realistic (physical) vacuum.
In particular, the whole $(M,\,\tan\beta)$ plane with $M<1000$~GeV is
disfavoured for these reason
in all the cases represented in fig.\,\ref{mtgb-iih}.
This is consistent with what was found for other superstring
and M-theory scenarios \cite{ibarra,ckm08}.

In the analysis above 
we have only considered the negative sign for the $\mu$
parameter. If $\mu>0$ was taken the experimental constraint on the
branching ratio of $\bsg$ would become significantly more
stringent, excluding regions of the parameter space with
$M<1000$~GeV for $\tan\beta\gsim30$. This rules out all the regions
where viable neutralino dark matter is found. Moreover, $\mu>0$ leads
to a negative SUSY contribution to the muon anomalous magnetic moment,
hence being further disfavoured.

\subsection{ B parameter constraint on the (I-I-I)-(I-I-A)
  configuration}

So far we have not imposed the boundary condition on the value of the
$B$ parameter at the GUT scale, which was also predicted in terms of
the modular weights and related to the rest of the soft parameters by
eq.(4.3).
As we already explained, in this approach the REWSB condition
(\ref{Bterm}) must be used in order to determine the value of
$\tan\beta$ by means of an iterative procedure. This leaves only one
free parameter, $M$, to describe all the soft terms and, if a solution
for the REWSB equations is found, a value of $\tan\beta$ is
predicted for each $M$.

\begin{figure}[t!]
    \hspace*{-0.5cm}
    \epsfig{file=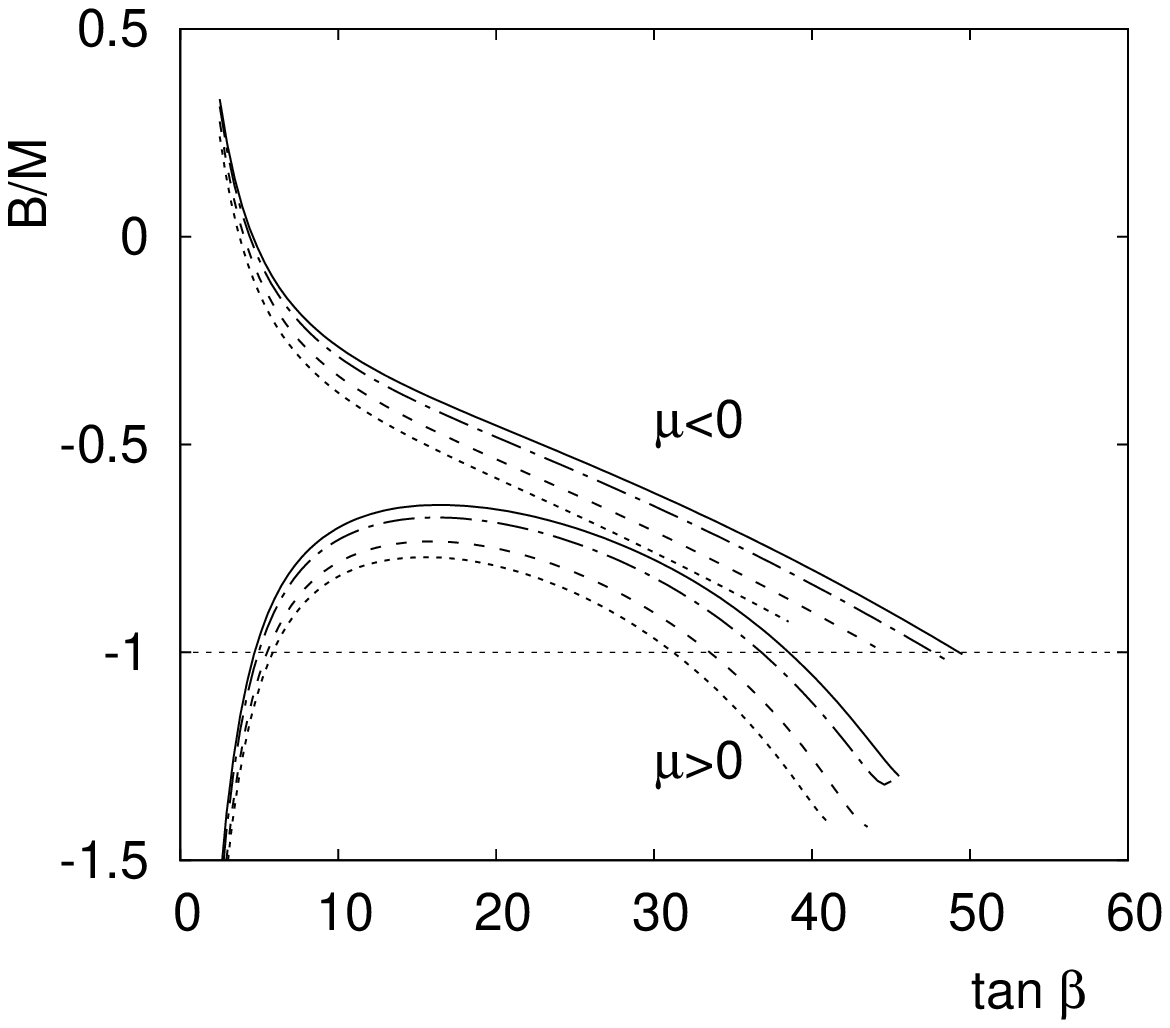,width=8cm}
    \epsfig{file=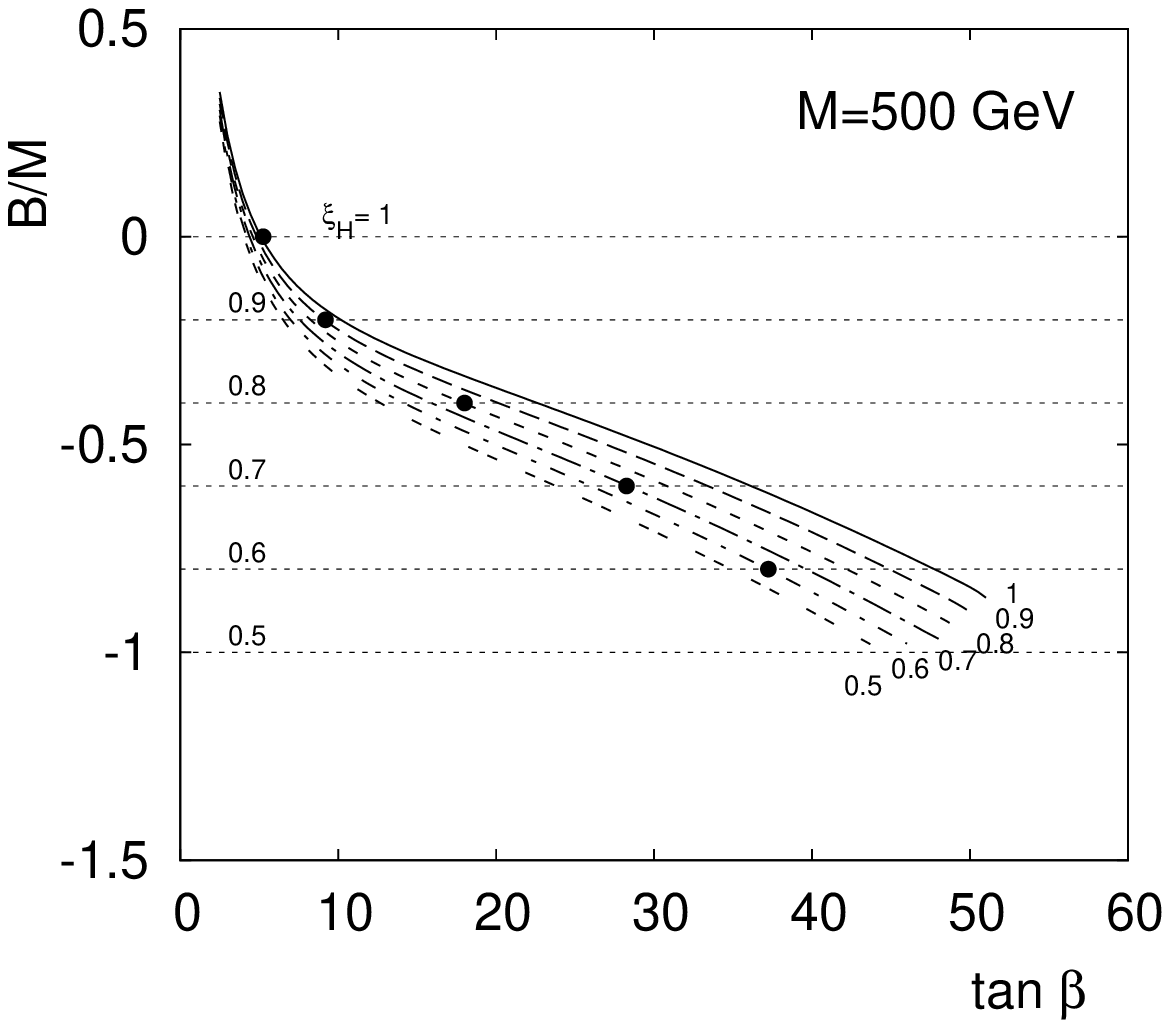,width=8cm}
  \caption{Left) Resulting $B(M_{GUT})/M$ as a function of
    $\tan\beta$ for the case with $\xi_H=0.5$. 
    The dotted, dashed, dot-dashed, and solid lines
    correspond to $M=300,\,500,\,1000$, and $1500\,{\rm
    GeV}$, respectively, for both
    signs of the $\mu$ parameter. 
    The boundary condition $B=-M$ 
    is indicated with a horizontal dotted
    line.  
    Right) The same, but for 
    the cases with $\xi_H=0.5,\,0.6,\,0.7,\,0.8,\,0.9$, and $1$, from
    bottom to top, with $\mu<0$ and $M=1000\,{\rm GeV}$.
    The corresponding boundary conditions for $B$ are represented with
    horizontal dotted lines, and 
    the solid circles indicate the values of $\tan\beta$ consistent
    with these. 
  }
  \label{btgb-iih}
\end{figure}

In fig.\,\ref{btgb-iih} (left panel) we display the value of $B/M$ at the GUT scale
as a function of $\tan\beta$ for several values of $M$ and for the two
possible choices of the sign of the $\mu$ parameter. 
The solutions of the REWSB equations correspond to the values of
$\tan\beta$ where the different lines intersect the dotted line, which
represents the boundary condition $B/M=-1$ in the $\xi_H=1/2$ case.
As we can see,  solutions are found for $\mu<0$ only when $\tan\beta$ is
very large.
In the $\mu>0$ case (disfavoured by $b\rightarrow s\gamma$ limits)
  solutions
are found  for $\tan\beta\sim4$ and  $30\lsim\tan\beta\lsim40$.

When the modular weight for the Higgses increases, the boundary
condition for $B$ is seriously affected. For instance, when 
$\xi_H=1$ as in case (I-I-A), one obtains $B=0$. 
As a consequence, the ranges of $\tan\beta$
which are solutions of the REWSB conditions change significantly. 
This is illustrated on the right hand-side of fig.\,\ref{btgb-iih},
where $B/M$ at the GUT scale is represented as a function of
$\tan\beta$ (with $\mu<0$ and $M=500$ GeV) 
for the cases with $\xi_H=0.5,\,0.6,\,0.7,\,0.8,\,0.9$,
and $1$, from bottom to top, respectively. 
The boundary conditions corresponding to these values of the Higgs
modular weights are represented by means of dotted lines, also from
bottom ($\xi_H=0.5$) to top ($\xi_H=1$).  
The solutions for $\tan\beta$ for each choice of modular weight
correspond to the intersection of the $B/M$ line with the
corresponding boundary condition, and are indicated with filled
circles.  
As we see in the figure, already slightly above $\xi_H=1/2$ (e.g. for $\xi_H=0.52$) 
correct EW symmetry breaking is obtained with the predicted $B$-term. This happens 
around tan$\beta\simeq 40$. As $\xi_H$ increases from $1/2$ to $1$ correct EW
symmetry breaking is obtained at the predicted $B$ with lower and lower
values of $\tan\beta$. In the (I-I-A) limit with $\xi_H=1$ solutions for
REWSB are  obtained for $\tan\beta\simeq 4$.



The resulting values for $\tan\beta$ as a function of the common scale
$M$ have been also superimposed on fig.\,\ref{mtgb-iih} by means of a thick
dot-dashed line. Consistently with what we just explained, 
when $\xi_H=1/2$ solutions are
only found for $\tan\beta\approx45$ and $M\gsim900\,{\rm GeV}$, in the
region with stau LSP. However, 
even with a slight increase in $\xi_H$,
 due to the rapid change in the
boundary condition for $B$, solutions of the REWSB
equations are found with smaller values of $\tan\beta$. 
If we want to obtain successful REWSB in a region
consistent with appropriate neutralino dark matter, one is lead only
to the region with $\xi_H\simeq 0.6$, quite close to the boundary
conditions (I-I-I) with $\xi_f=\xi_H=1/2$.
This may be seen in fig.(\ref{mtgb-iih})
(upper right)
in which the thick dot-dashed line is very close to the line
marking the separation between neutralino and stau LSP regions,
i.e., the coannihilation region.
On the other hand, in the  (I-I-A) case with $\xi_H\simeq 1$ one obtains
appropriate REWSB for tan$\beta\simeq 4$, far away from the coannihilation 
region and hence too much dark matter is predicted.
Thus  insisting in getting neutralino dark matter consistent 
with WMAP measurements and consistent REWSB selects the region 
close to the  (I-I-I) boundary conditions in which all 
MSSM matter fields live at intersecting 7-branes.

It is worth mentioning here that variations in the value of the top
mass slightly alter the running of the $B$ parameter and, as a
consequence, lead to a small shift in the solutions for
$\tan\beta$. We have checked that in the previous examples this shift
is $\Delta\tan\beta\approx\pm1$ when the top mass varies from 
$m_t=169.2\,{\rm GeV}$ to $m_t=174.8\,{\rm GeV}$ (which corresponds to
a $2\sigma$ deviation from the experimental central value). 
Although smaller top mass also implies a more stringent constraint
from the experimental bound on the lightest Higgs mass, the rest of
the constraints are not significantly affected and 
the coannihilation region remains basically unaltered. One can
therefore understand $\Delta\tan\beta$ as a small uncertainty on the 
trajectories for $\tan\beta$ in 
fig.\,\ref{mtgb-iih} to be taken into account when demanding
compatibility with the regions with viable neutralino dark matter.

\subsection{The bulk 7-brane (A-A-$\phi$) configuration}

Let us consider now the other alternative left, in which the MSSM resides
at the bulk of the 7-branes. As seen in table 1,
in this case  the sfermion soft masses vanish at the GUT scale. This
has important implications on the resulting low-energy
spectrum. Although squark masses (which receive large positive
contributions in the corresponding RGEs from the gluino mass
parameter) easily become large enough, slepton masses remain rather
light. This is particularly problematic for the lightest stau, 
due to the negative contribution proportional to the Yukawa in the RGEs. 
For this reason the stau mass-squared becomes negative even for
moderate values of $\tan\beta$. In this example, this sets an upper
bound of $\tan\beta\lsim25$.

\begin{figure}[t!]
  \epsfig{file=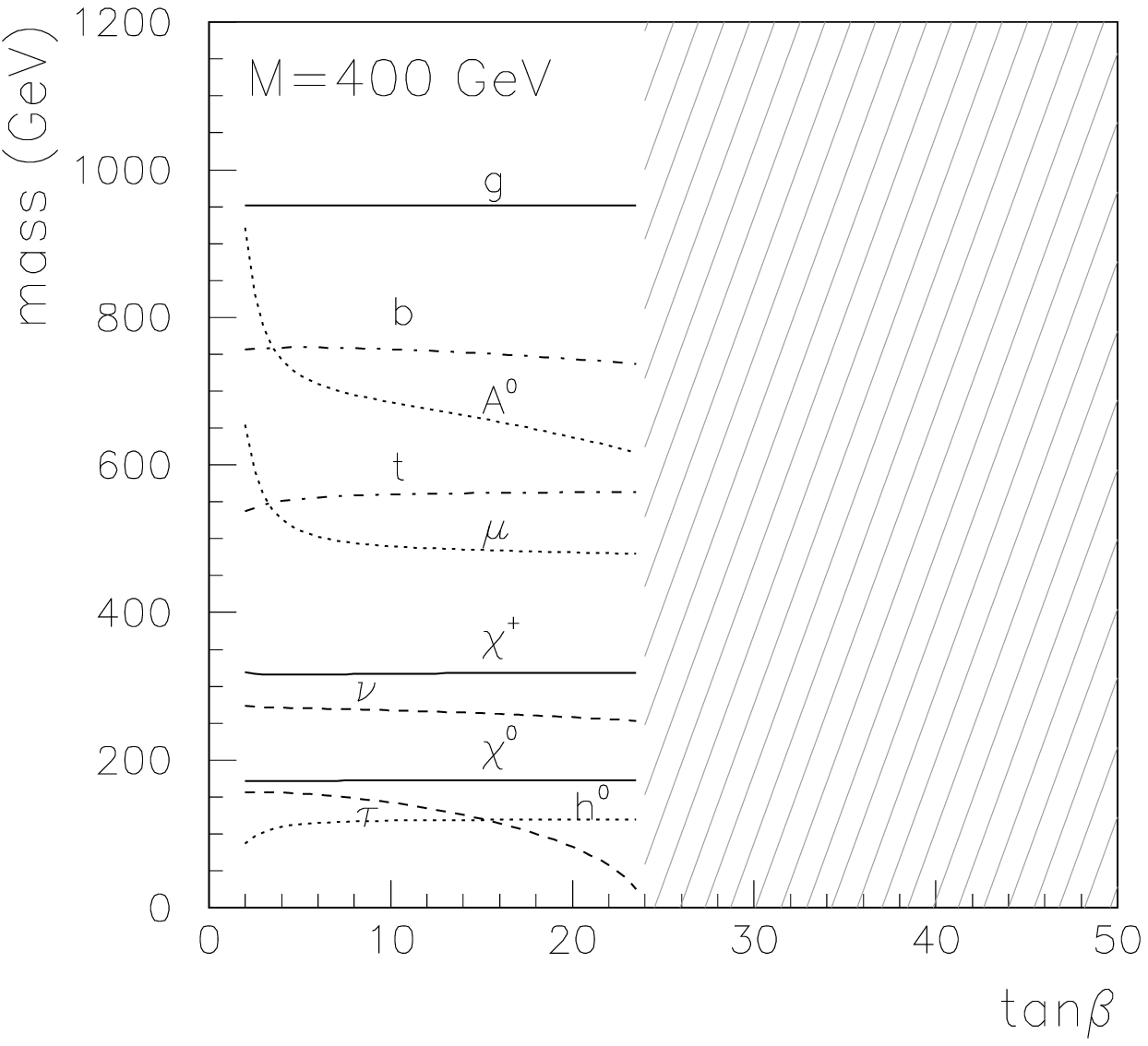,width=8cm}
  \hspace*{-1cm}
  \raisebox{2ex}{\epsfig{file=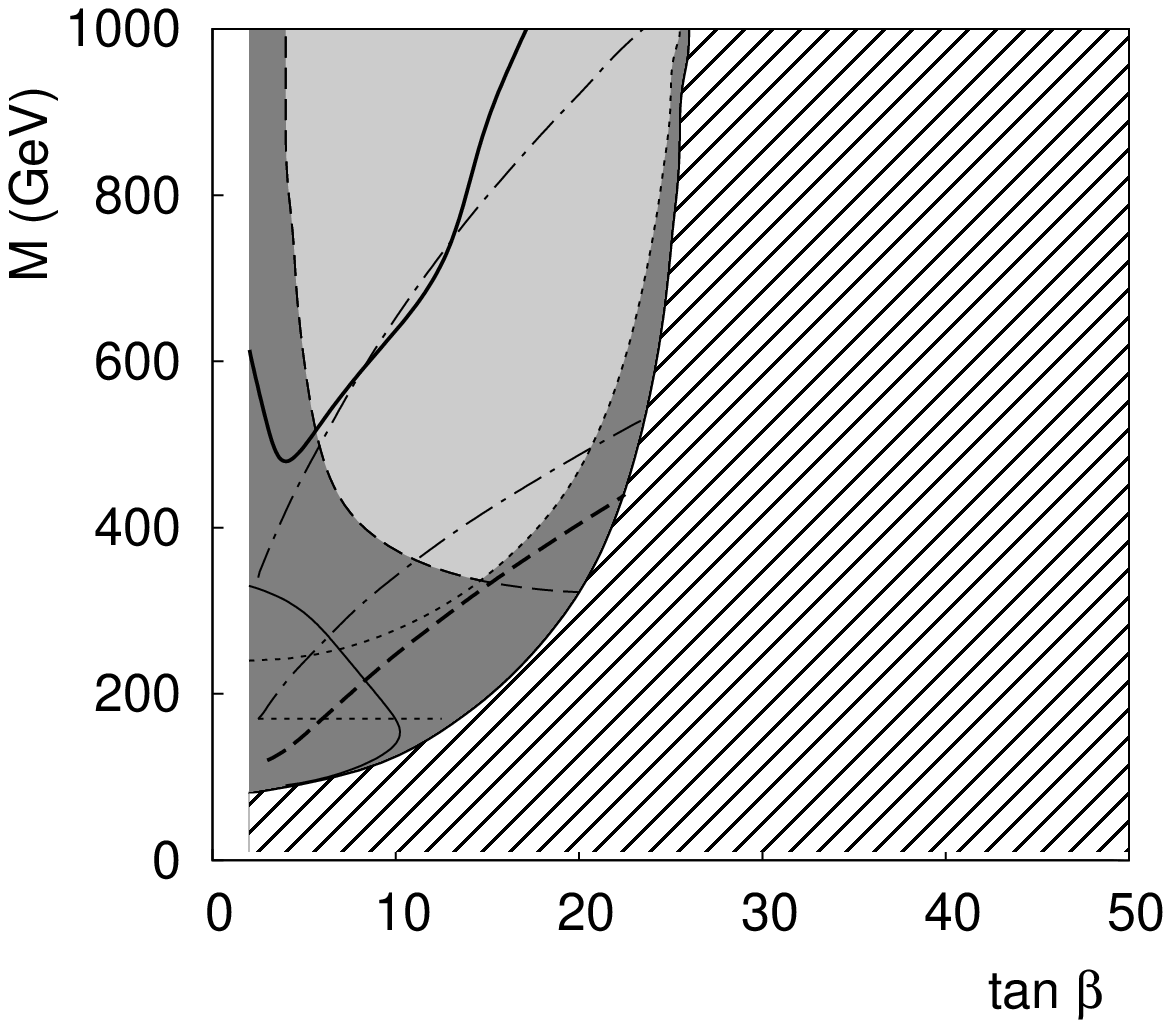,width=8cm}}
  \caption{Left) Low-energy supersymmetric spectrum as a function of
  $\tan\beta$ for case (A-A-$\phi$). 
    Right) Effect of the various experimental constraints on the
    $(M,\,\tgb)$ plane for case (A-A-$\phi$). 
    Colour and line conventions are the same as in
    fig.\ref{mtgb-iih}.
    In addition, the UFB constraints are fulfilled in the region above
    the thick solid line. 
    }
  \label{mtgb-aap}
\end{figure}

In the remaining allowed areas of
the parameter space the stau becomes the LSP. A stable 
charged LSP
would bind to nuclei, forming exotic isotopes on which
strong experimental bounds exist. Phenomenological consistency would
then require such staus to decay, a possibility which arises it
R-parity was broken. There is therefore no viable supersymmetric dark
matter candidate in this scenario.

The resulting SUSY spectrum, together with the effect of the rest of the
experimental constraints on the parameter space are shown in
fig.\,\ref{mtgb-aap}, clearly displaying all the above mentioned
features. 
Interestingly, and contrary to what we observed for the intersecting
$7$-brane configurations, there is a region of the parameter space which
satisfies the UFB constraints, corresponding to the area above the
thick solid line with $M\gsim 500$~GeV and $\tan\beta\lsim 20$. 
This is possible because of the increase in the Higgs mass parameters
at the GUT scale (remember that now $m_{H_u,H_d}^2=M^2$), which entails
a less negative contribution to the
scalar potential along the UFB directions.

If we further impose the prediction for the 
B-parameter the situation is worse. Indeed 
 in this scenario the boundary condition for $B$ at the
string scale is $B=-2M$. No solutions are found for $\tan\beta$
neither for $\mu<0$, nor for $\mu>0$ satisfying this boundary condition.


\subsection{Effect of magnetic fluxes}

We have seen how the intersecting 7-brane configuration (I-I-I)-(I-I-A) 
with $\xi_H\simeq 0.6$  is consistent with all constraints 
including appropriate amount of neutralino dark matter. 
This 'effective modular weight' $\xi_H\simeq 0.6$ could be understood
if the Higgs field is a linear combination of fields with 
modular weights $\xi_H=1/2$ (predominant) and $\xi_H=1$.

Alternatively one could think that there could be some
higher order correction which could explain the small deviation from
the fully intersecting configuration (I-I-I) with $\xi_H=1/2$.
As we discussed, one possible source for such small corrections could
be the magnetic fluxes which are anyway required for the 
spectrum to be chiral. Using the results in section (3.2) we can estimate
what could be the structure of such corrections.
In agreement with the unification hypothesis, 
we will assume that all sfermions have  
the same flux correction in their  Kahler metrics proportional to some 
parameter $c_f$. We will then parametrize the corrections in terms of
parameters defined for the different cases as:
\beq
\rho=\frac {(c_H-as)}{t} \ ;\ 
\sigma=\frac {as}{t} \ ;\ \rho_f=\frac {c_f}{t^{1/2}} \ ;\ 
\rho_H=\frac {c_H}{t^{1/2}}\, ,
\eeq
where $a$ is defined in eq.(\ref{metricaaprox}) (in our case $a_i=a$
for gauge coupling unification). The results for the soft terms are
shown in table 2. 
\begin{table}[htb] \footnotesize
\renewcommand{\arraystretch}{1.50}
\begin{center}
\begin{tabular}{|c||c|c|c|c|}
\hline   Coupling &
    $m_f^2$
 &   $m_H^2$
 &   A     &  $B$        \\
\hline\hline
   (A-A-$\phi$) &
    0       &   $ {|M|^2(1-2\rho)}$ &   $-M(1-\rho )$   &  $-2M(1-\rho)$  
  \\
\hline
    (I-I-A)  &
   $\frac {|M|^2}{2}(1-\frac {3}{2}\rho_f)$ 
   &    0    &  $-M(1-\rho_f)$  &  0     \\
\hline
   (I-I-I)  &
   $\frac {|M|^2}{2}(1-\frac {3}{2}\rho_f) $  &   $\frac {|M|^2}{2}(1-\frac {3}{2}\rho_H)$    
&  $-\frac {1}{2}M(3-\rho_H-2\rho_f)$    &  
$-M(1-\rho_H)$  
\\
\hline \end{tabular}
\end{center} \caption{\small  Corrections from magnetic fluxes to the 
different soft terms. The parameters $\rho,\rho_H, \rho_f$
are defined in the main text. }
\label{correcciones de flujos}
\end{table}
To get the results we have assumed that
$\rho_H,\rho_f \gg \sigma$, given their different large $t$
behaviour. Looking at the table one observes as a  general  conclusion
that the size of scalar soft terms decreases with respect to gaugino
masses as fluxes are turned on. This is consistent with the known fact
that as fluxes increase 7-branes localize and get boundary conditions
more and more similar to 3-branes, whose  matter fields  are known do
not get  bosonic soft terms.

Note  that, as we mentioned, the scalar masses of fields of A-type
(modular weight $\xi =1$) remain massless even after the addition of
magnetic fluxes. So the problem of the scheme (A-A-$\phi$)
of having too light (or even tachyonic) right-handed sleptons
cannot be solved with the addition of fluxes.

Interestingly, the inclusion of magnetic fluxes also
alters the boundary condition for the bilinear parameter. 
In particular, for case (I-I-I), 
$B$ at the GUT scale becomes less negative. This is welcome, as we
already saw in the previous subsection, in order to obtain successful
radiative electroweak symmetry breaking.

\begin{figure}[t!]
  \hspace*{-0.6cm}
  \epsfig{file=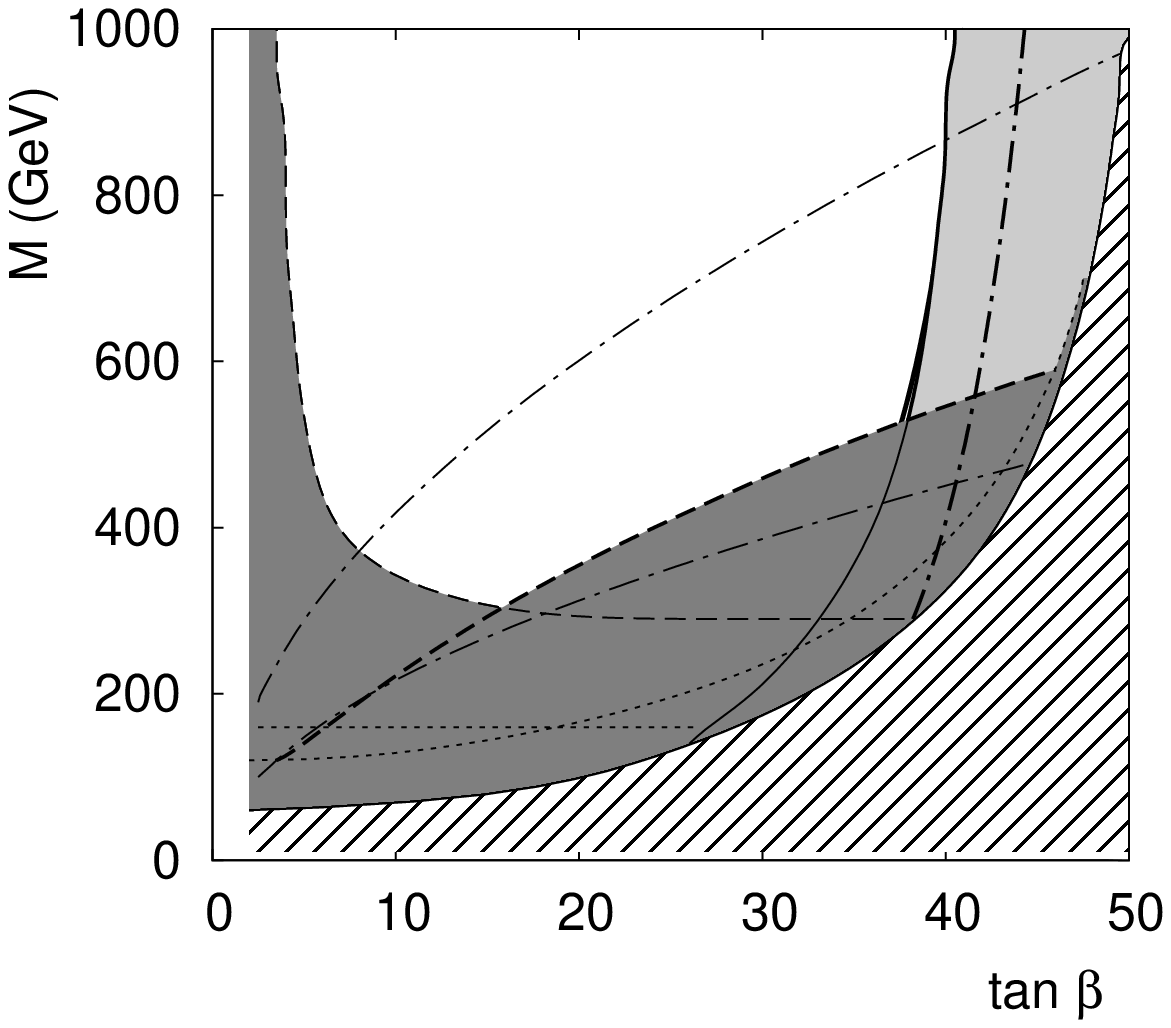,width=8.cm}
  \epsfig{file=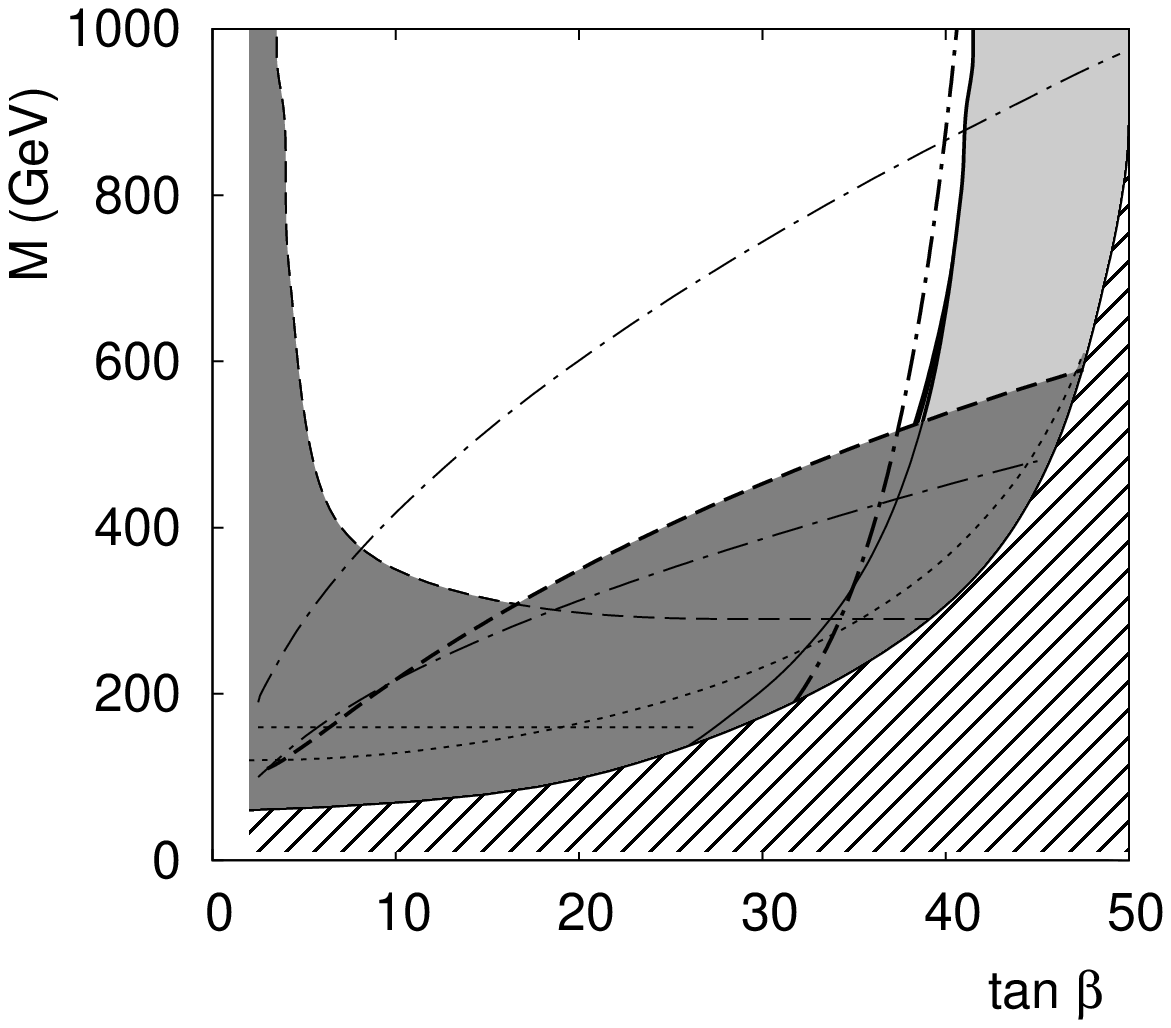,width=8.cm}
  \caption{The same as in fig.\,\ref{mtgb-iih} but for case (I-I-I)
  where corrections coming from fluxes are included with $\rho_H=0.1$
  (left) and $\rho_H=0.2$
  (right). In both of them $\rho_f=0$.
    }
  \label{mtgb-iiiflux}
\end{figure}

We have explored the effect of a small flux correction 
to the (I-I-I) case ($\xi_H=1/2$)  with
$\rho_H=0.1,0.2$ and $\rho_f=0$.
The results are shown in fig.\,\ref{mtgb-iiiflux}. The low-energy
supersymmetric spectrum is not very affected by the change on the
Higgs mass parameters. The only visible effect is a very slight
increase in the stau mass for large $\rho_H$ as a consequence of the
decrease in $|A_L|$. Hence, the allowed region with neutralino LSP and
correct dark matter abundance barely changes. 
On the contrary, 
the line along which the boundary condition for $B$ is satisfied
changes significantly, being shifted towards smaller $\tan\beta$ as
$\rho_H$ increases. Compatibility with viable neutralino dark matter
is found around $\rho_H\approx0.2$ with $\xi_H=1/2$.
Thus indeed, magnetic fluxes could provide for the small correction
required to get full agreement with the appropriate dark matter
for the (I-I-I) intersecting 7-branes scheme.

Note that 
the corrections to sfermion masses, parametrized by $\rho_f$, imply a
decrease of their mass at the GUT scale. This leads to lighter staus
at the EW scale, and therefore the region where the neutralino is the
LSP (and obviously the region with correct relic density) is shifted
towards smaller values of $\tan\beta$. This would make it more
complicated to obtain compatibility of successful REWSB and neutralino
dark matter (larger $\rho_H$ would be needed).

\section{Implications for LHC}

In the previous subsections we have seen how case 
(I-I-I) is singled out as the one in which REWSB can be
reconciled with experimental and astrophysical constraints more
easily, either via a small mixing of the Higgs or by the inclusion of
small magnetic flux corrections.
Indeed all constraints are fulfilled by soft term boundary conditions
\begin{eqnarray}
        m_f^2&=& 1/2 \,|M|^2\,,\nonumber\\
        m_H^2&\approx& (1/2\, -\, 0.1)\,|M|^2\,,\nonumber\\
        A_{U,D,L}&\approx&(-3/2\, +\ 0.1)\,M\,,\nonumber\\
        B&\approx& -(1\, -\, 0.2) \,M\,.
        \label{soft-optim}
\end{eqnarray}
remarkably close to the values obtained in a 
 configuration  (I-I-I) in which 
all quark, lepton and Higgs fields lie at intersecting 7-branes.
It is then natural to wonder whether this whole class of string
scenarios can be tested at the LHC and what might a typical signal
look like. 
In order to estimate the potential LHC reach on the
$(M,\,\tan\beta)$ planes previously shown we have followed the same
approach that was used in \cite{eos08} to explore the discovery
potential of various SUSY scenarios.

Namely, we assume the canonical signature for SUSY searches,
consisting in the observation of missing transverse
energy and multijet events. These effects can be produced in the
decays of squarks and/or gluinos into a final state with 
a pair of neutralinos plus the corresponding hadronic products. The
expected  sensitivity of the CMS detector to this kind of signal 
was explored in \cite{cms07}, where the projected $5\sigma$ reach
contours were extracted for $1$~fb$^{-1}$ and $10$~fb$^{-1}$ 
in the ($M,\,m$) plane of the Constrained
Minimal Supersymmetric Standard Model (CMSSM), where soft parameters
are considered to be universal at the GUT scale.

The sensitivity to this kind of signal is mostly
dependent on the gluino and squark masses, since these determine their
production cross section. For this reason, one can
express the former reach contours as functions of $m_{\tilde g}$ and
$m_{\tilde q}$ and thus derive the corresponding contours in the 
($M_{\tilde g},\,m_{\tilde
q}$) plane. In doing so, and following the procedure of \cite{eos08}, 
we have used a numerical fit to approximate the contours of fig.\,13.5
of \cite{cms07}, then using the relations of the gluino and squark
masses with the $M$ and $m$ parameters of the CMSSM.

This is very useful, because it allows us to estimate the potential
LHC reach on scenarios beyond the CMSSM, such as the ones obtained
in this work and express them in terms of reach contours on 
the $(M,\,\tan\beta)$ plane.
In fact, the soft terms in eq.(\ref{soft-optim})
do not display a large departure from the universal structure 
of the CMSSM and 
would correspond to the region with a small
common scalar mass.
Interestingly, 
these are conditions under which the resulting squarks are
lighter than the gluino, which
leads to large rates in the production
processes of $\tilde q\tilde q$, $\tilde q\tilde g$ and $\tilde
g\tilde g$ (see, e.g., \cite{bbbkt03}). Consequently, this region of
the parameter space is more easily explored through searches for 
missing transverse energy.

The resulting reach contours show in our case 
no dependence with $\tan\beta$ 
(this is consistent with the CMSSM, for which 
the reach contours for this particular signal were also
found to be mostly insensitive to $\tan\beta$, see
e.g., \cite{bbbkt03}). 
Hence, the reach contours on 
the $(M,\,\tan\beta)$ plane are simply horizontal lines for a given
value of the common gaugino mass, $M$. More specifically, the region
of the parameter that could be explored by CMS with a luminosity of
$1$~fb$^{-1}$ corresponds to $M\lsim650$~GeV in all the plots of
fig.\,\ref{mtgb-iih}. Notice that this is enough to start probing
some of the regions where consistent REWSB can be obtained while
fulfilling all experimental constraints and having viable neutralino
dark matter. In these areas the gluino and squark masses are 
$m_{\tilde g}\lsim 1.5$~TeV and
$m_{\tilde q}\lsim 1.3$~TeV, corresponding also to neutralinos with a
mass of $m_{\tilde\chi^0}\lsim300$~GeV. 
With a luminosity of $10$~fb$^{-1}$, LHC would be able to explore the
whole region of the parameter space with $M\lsim900$~GeV. This
corresponds to gluinos with a mass $m_{\tilde g}\lsim 2$~TeV and
squarks with $m_{\tilde q}\lsim 1.8$~TeV. The neutralino mass is
$m_{\tilde\chi^0}\lsim400$~GeV in this region.

Due to the constraint on the neutralino relic density, 
the supersymmetric spectrum is characteristic of the CMSSM in the
coannihilation region, featuring a very small mass difference between
the lightest neutralino and the lightest stau. A typical signal in
this region would consist in looking for missing energy in 
the decay chain of the second-lightest neutralino into the lightest
one,
$\tilde\chi^0_2\to\tau\tilde\tau_1\to\tau\tau\tilde\chi^0_1$. In this
case, end-point measurements at the LHC could make it possible to
determine the neutralino-stau mass difference 
\cite{coan-lhc}.

\section{Final comments and conclusions}

With the advent of LHC the possibility exists of SUSY particles 
being produced and their masses being measured. 
If indeed SUSY particles are found and (at least some of) their masses 
are measured we will have to address the issue of the origin 
of SUSY-breaking. We have argued in the present article that 
such measurements could provide important information 
which could rule out (or favour) large classes of 
MSSM-like possible string compactifications. 

Modulus dominated SUSY breaking is singled out in string theory 
as a way to obtain SUSY-breaking vacua at the classical level with an
approximately vanishing cosmological constant. In Type IIB orientifold
compactifications it corresponds to the presence of certain
class of SUSY breaking antisymmetric fluxes, a degree of freedom which
is generically present in such compactifications. 
We have addressed in this paper what could be the possible patterns 
of SUSY breaking masses under the assumption of modulus dominance 
in string theory. We have argued that this assumption together with other
phenomenologically motivated ones (MSSM spectrum, existence of one large (top) Yukawa coupling and 
tree level gaugino masses) identify generic classes of possible string
compactifications. These are Type IIB orientifold compactifications in
which the MSSM particles reside either in the bulk or at intersections of 7-branes. 
If one further insists in a unification of coupling constants that points 
in the direction of F-theory 7-branes  in which underlying unification
structures like $SO(10)$ are possible.

In order to compute the effective action and SUSY breaking soft terms 
the dependence of the metrics of MSSM matter fields on the 
Kahler moduli are required.  Those matter metrics  may be described 
for the case of a single local Kahler modulus $t$  in terms of 
'modular weights' $\xi_\alpha $. We have argued that both for simple toroidal 
settings as well as more complicated ones like 'swiss cheese' CY  or
local F-theory configurations there are  three classes of 
matter fields I,A,$\phi$ with modular weights $\xi=1/2,1,0$, 
respectively.

The possible modular weights of MSSM fields
are further constrained by the condition that at least one
large Yukawa coupling (that of the top) should exist.
It turns out that non-vanishing trilinear Yukawa couplings in this setting 
only exist for combinations of type 
 (I-I-I),(I-I-A) and (A-A-$\phi$).
In the first two cases MSSM fermions reside at intersections (I) of 
the 7-branes whereas the Higgs multiplet resides either at an
intersection (I) or in the bulk (A) of a 7-brane. In the last case
all MSSM particles live in the bulk of a stack of 7-branes 
but come from zero modes of 8-dimensional vector multiplets (A)
or scalar multiplets ($\phi$). 

With the knowledge of the modular weights and assuming that 
SUSY-breaking is dominated by the auxiliary field of the local
Kahler modulus $t$ one can explicitly compute the corresponding
SUSY-breaking soft terms for the three configurations, which 
are given in table 1. We have also estimated subleading 
corrections coming from  magnetic fluxes in the bulk 
of 7-branes which are in general required to get chirality.

Taking these soft terms as boundary conditions at the string/GUT scale, 
we have computed the low energy SUSY spectra by numerically solving 
the renormalization group equations for soft terms, imposing
standard radiative electroweak symmetry breaking. We have performed
this computation for the three 7-brane configurations allowing for
couplings (I-I-I), (I-I-A) and (A-A-$\phi$). 
The scheme is very predictive, there are
only two parameters, the gaugino mass $M$ and $\tan\beta$ which 
is taken  as a free parameter. Once radiative EW symmetry breaking
is imposed, essentially only one free parameter is left,  $M$,
which sets the overall scale of soft terms.

We impose consistency of the resulting supersymmetric spectrum 
with the most recent experimental and
astrophysical constraints. Namely, we include bounds on the 
Higgs and sparticle masses, as well as on low-energy observables such as 
the branching ratios of 
rare decays ($b\rightarrow s\gamma$, $\bmumu$) and the
supersymmetric contribution to the muon anomalous magnetic moment,
$(g-2)_\mu$.  
Moreover, we have also studied the viability of neutralino dark matter
by requiring its relic density to fulfil the WMAP bound on the
abundance of cold dark matter. 
All these constraints essentially single out 
the configuration (I-I-I) in which all MSSM chiral fields reside at
the intersection of 7-branes. Specifically,  the modular weights required
to pass all experimental and dark matter constraints are given in 
eq.(\ref{soft-optim}) and are very close to those corresponding 
to universal modular weight $\xi=1/2$ for all MSSM chiral fields.
We have argued  that 
the small deviations of eq.(\ref{soft-optim}) from those universal
boundary conditions may be understood as coming from small effects,
such as the presence of magnetic fluxes.

It must be emphasized that the fact that this (I-I-I) option 
passes all phenomenological tests is quite non-trivial, since, as we said, 
after imposing radiative electroweak symmetry breaking
there is essentially only one free parameter, $M$. Furthermore the scheme
is quite predictive and could be tested at LHC if SUSY particles are found.
The overall structure of the SUSY spectrum is shown in fig.\,\ref{spectrum-iih}
(left figure for tan$\beta\simeq 35$). One of the prominent properties is that a
charged stau will be only slightly heavier than the lightest neutralino.
This is to be expected because appropriate dark matter here appears 
because of a coannihilation effect.  
We have shown that, with a luminosity of
1 fb$^{-1}$,  LHC should be able test the present scheme for $M<600$
GeV, whereas
$M<900$ GeV could be reached for  10 fb$^{-1}$, using the missing energy
signature.  A more detailed study of possible signatures at LHC of the
present scheme would be quite interesting. We are looking forward to the
forthcoming LHC results. If sparticles are found we might be checking not
only SUSY but a large class of string theory compactifications.

\vspace*{2cm}

{\bf Acknowledgments}\\
We thank  A. Font, F. Marchesano, F. Quevedo, P. Slavich  and A. Uranga , for useful discussions.
This work  has been supported by the European
Commission under RTN European Programs MRTN-CT-2004-503369,
MRTN-CT-2004-005105, by the CICYT (Spain) under project
FPA2006-01105, the Comunidad de Madrid under project HEPHACOS
P-ESP-00346 and the Ingenio 2010 CONSOLIDER program CPAN.
L. Aparicio acknowledges finantial
support of 
the Ministerio de Educaci\'on y Ciencia of Spain through a FPU grant. 
D.G. Cerde\~no is supported by the program ``Juan de la Cierva'' of
the Ministerio de Educaci\'on y Ciencia of Spain, Ayudas de I+D
S-0505/ESP-0346, and by the EU research and training network
MRTN-CT-2006-035863.

\clearpage

\appendix

\section{Alternative sources for a $\mu$-term }

In the main text it is assumed that there is an explicit $\mu$-term,
possibly induced by fluxes.
This seems like the simplest possibility. However, for generality
 we summarize in this appendix other alternatives
or modifications concerning the $\mu$-term.
As we will see these other alternatives do not seem to be particularly
more promising concerning the generation of $\mu$ and $B$ parameters
consistent with radiative electroweak symmetry breaking.

An explicit $\mu$-term from fluxes may be  forbidden if the
Higgs bilinear is charged under some  (anomalous and massive)
$U(1)$.
In this case  a possible  source for the generation of $\mu$- and B-terms
could  be some sort of string theory non-perturbative effect.
As remarked in  \cite{instantons} string instantons may give rise to
Higgs bilinear superpotentials of the form
\beq
 \ M_s e^{-S_{ins}}\ H_uH_d \ =\ M_s e^{-\sum_id_iT_i} \ H_uH_d\, ,
\eeq
where the instanton action $S_{ins}$ is given by a particular
linear combination of Kahler moduli characteristic of the
contributing instanton. As we said,
in string compactifications the Higgs
bilinear may be  charged under some (anomalous and massive) extra $U(1)$
and the full term is rendered gauge invariant because the particular combination of
Kahler moduli in the exponent shifts under gauge transformations appropriately.
Such a generation of a $\mu$-term has the shortcoming that it does not
solve the $\mu$-problem: there is no reason why  $\mu$ and the soft terms
should end up being of the same order of magnitude. That problem
could be solved
if, instead of a superpotential $\mu$-term, a Giudice-Masiero $Z$
function is generated. In some heterotic orbifold compactifications
\cite{gmstrings} (and their Type I duals)
bilinear terms with the structure of a $Z$ term do actually appear with the
structure
\beq
   \frac {1}{(T_i+T_i^*)}   \phi_1 \phi_2 \ +\ h.c.\, ,
\label{insGM1t}
\eeq
where $\phi_{1,2}$ are chiral fields forming a a
vectorlike pair.  That happens in Abelian
heterotic compactifications with the $\phi_{1,2}$ fields
corresponding to untwisted matter of a complex plane
only affected  by an order two twist.
To our knowledge no concrete semirealistic model in which
one can identify $\phi_1,\phi_2$ fields with $H_u,H_d$ has
been constructed, but still it shows that in principle such
a GM term could be present at the tree level.
On the other hand in some  string compactifications the Higgs
bilinear $(H_uH_d)$ is charged under some extra (anomalous and
massive) $U(1)$ gauge symmetry so that such a tree level $Z$-term
is forbidden perturbatively by gauge invariance.
In such a case it is conceivable that string instanton effects
could generate a non-perturbative $Z$-term with a structure
\beq
e^{-S_ins} \   (K_{H_u}K_{H_d})^{1/2}\ H_uH_d \ +\ h.c. \ \ .
\label{insGM2}
\eeq
Again the gauge transformation under some extra $U(1)$ of the
Higgs bilinear would  be compensated by the transformation of
the instanton action $S_{ins}$. If such a term is present and the
exponential factor does not provide too much  supression,
then both a $\mu$-term and a $B$-term would be generated of the same
order of magnitud as the rest of the soft terms.

Let us discuss for completeness
what are the results for the $B$-parameter obtained
for these new cases. One could first consider
   that $\mu$ could have a
non-perturbative (e.g. instanton) origin with
non-perturbative
dependence on the local Kahler  parameter $t$, i.e
$\mu = ae^{-bT}$ with $T$ the local modulus and a,b constants (more generally
dependent on the complex structure fields, which we assume to be fixed
at their VEV). Under those
circumstances the  expression for $B$  changes slightly:
\beq
B\ =\ -M(2-\xi_{H_u}-\xi_{H_d}) \  -\ Mbt\, .
\eeq
Note that in order to have a physical $\widehat \mu$-term hierarchicaly smaller than
a string scale of order $M_{GUT}$ one should have  $bt\propto \log(M_s/{\tilde \mu})
\propto 30$. This would give rise to a $B$-term much larger than the rest of the
soft terms and would not be viable phenomenologically. Thus this possibility does
not look very promising.

As we said, another option is the presence of
a Giudice-Masiero term  present
already at the perturbative level.
 We could parametrize it, in analogy with the diagonal
Higgs Kahler metric:
\beq
Z\ =\ \frac {t^{(1-\xi_Z) }}{t_b}  \ \ ;\ \ \xi_Z=\frac {(\xi_{H_u}+\xi_{H_d})}{2}\, .
\label{estazeta}
\eeq
Such a GM  term  would give rise to
 result for the $\mu $ and B-terms:
\beq
{\widehat {\mu} }\ =\  -M(1- \xi_Z) \ \  ;\ \
B_Z \  =  \ -M (2\ \ -\xi_Z)\, .
\eeq
Note that for $\xi_Z=1$ (corresponding to a (I-I-A) scheme with $\xi_H=1$)
no $\mu$-term  is generated so in this case the GM mechanism does not
provide for the required term. On the other hand, for $\xi_Z=1/2$,
$\mu=-M/2$ is generated. In our numerical solution of the RGE we have found that
the required $\widehat {\mu}$ is always substantially larger than $M$, so that
we would not get appropriate radiative symmetry breaking for $M/2$  either.
All in all, a GM mechanism does not seem sufficiently flexible to
be compatible with radiative symmetry breaking in the present scheme.

In the presence of magnetic fluxes
one expects some corrections to these results.
In particular
 if in analogy to the diagonal
Higgs metric one assumes a Giudice-Masiero term given by
$Z=(t^{1-\xi_Z}+ c_Z)/t_b$  with  $\xi_Z=\xi_{H_u}=\xi_{H_d}=\xi_{H}$
and $c_Z=c_{H_u}=c_{H_d}=c_H$ one
obtains
\beq
\mu\ =\ -\frac{F_t}{t} (1-\xi_H)(1-\frac {c_H}{t^{1-\xi_H}})
\ \ ;\ \
B_Z\ =\ -\frac {F_t}{t}
(2-\xi_H(1-\frac {c_H}{t^{1-\xi_H}}))\, .
\eeq
Still in this case the generated ${\widehat \mu}$-term can only
be smaller so that again it will not be consistent with
appropriate radiative EW symmetry breaking.

One could finally consider the possible presence of a
GM term as in eq.(\ref{estazeta}) appearing at the non-perturbative
level from string instanton effects:
\beq
 Z\ =\ (\alpha e^{-\beta t})\ \frac {t^{(1-\xi_Z) }}{t_b}
\label{insGM1}\, ,
\eeq
with $\alpha,\beta$ $t$-independent quantities.
With this ansatz one finds for the $B$ and $\widehat \mu$ parameters
\beq
{\widehat {\mu}} \ =\  -M\alpha e^{-\beta t}(1-  \xi_Z+\beta t) \ \  ;\ \
B_Z \  =  \ -M \frac {(2 -\xi_Z)(\xi_Z-1)+\beta^2t^2}
{(\xi_Z-1)-\beta t}\, .
\eeq
Unlike the case of an explicit perturbative $T$-dependent $\mu$-term, here $\beta t$
needs not be large, and (asuming the exponential factor does not provide for
a large supression) the $\mu$ term would be naturally of order the SUSY breaking
soft terms. Still, within the spirit of the present paper in which the Kahler moduli are taken to be
large, one expects in general non-perturbative correction to be rather small.

\end{document}